\documentclass[11pt]{amsart}
\usepackage{amssymb}                    
\usepackage{graphicx}

\usepackage[centertags]{amsmath}
\usepackage{amsfonts,amssymb}
\usepackage{times}
\usepackage{subfigure}
\usepackage{cite}
\usepackage{verbatim}

\newtheorem{proposition}{Proposition}

\newcommand{\NN}{\mathbb{N}}
\newcommand{\RR}{\mathbb{R}}

\def\lj38{{\rm LJ}_{38}}

\def\<{\langle}
\def\>{\rangle}

\begin{document}


\title[Flows in Complex Networks]{Flows in Complex Networks: Theory,
  Algorithms, and Application to Lennard-Jones Cluster Rearrangement
}


\author{Maria Cameron} 
\address{University of Maryland, Department of
  Mathematics, College Park, MD 20742} 
\email{cameron@math.umd.edu}
\author{Eric Vanden-Eijnden} 
\address{Courant Institute of
  Mathematical Sciences, New York University, 251 Mercer Street, New
  York, NY 10012}
\email{eve2@cims.nyu.edu} 


\maketitle

\begin{abstract}  
  A set of analytical and computational tools based on transition path
  theory (TPT) is proposed to analyze flows in complex networks.
  Specifically, TPT is used to study the statistical properties of the
  reactive trajectories by which transitions occur between specific
  groups of nodes on the network.  Sampling tools are built upon the
  outputs of TPT that allow to generate these reactive trajectories
  directly, or even transition paths that travel from one group of
  nodes to the other without making any detour and carry the same
  probability current as the reactive trajectories. These objects
  permit to characterize the mechanism of the transitions, for example
  by quantifying the width of the tubes by which these transitions
  occur, the location and distribution of their dynamical bottlenecks,
  etc.  These tools are applied to a network modeling the dynamics of
  the Lennard-Jones cluster with 38 atoms ($\lj38$) and used to
  understand the mechanism by which this cluster rearranges itself
  between its two most likely states at various
  temperatures.\keywords{transition path theory \and self-assembly
    \and protein folding \and glassy dynamics \and Markov State
    Models}
\end{abstract}

\section{Introduction}
\label{sec:intro}

In recent years, networks have gained popularity as a tool to
represent, organize, and interpret phenomena arising in many fields of
science, including physics, biology, social sciences, etc. Questions
as diverse as the structure of the World Wide Web, the robustness of a
nation's banking system or its power grid, or the mechanism of
functions inside a cell can be expressed in terms of networks.  These
applications have led to networks whose structure and complexity have
gone far beyond the examples studied before in the classical computer
science literature. Driven partly by the emergence of these new
applications, research in network science has also undergone a
revolutionary change in recent years. While traditional network
science was basically a subject of graph theory and focused on
networks with rather simple structure, recent studies often took the
viewpoint of treating networks as complex systems, and used tools and
concepts from statistical mechanics. While the structure and topology
of networks has been under much investigation, the dynamics on the
network is less well understood despite the fact that it leads to
important and nontrivial questions. For example, any network with
positive-weighted edges defines a Markov jump process (MJP) (and vice
versa) and in many applications, it is of interest to understand the
interplay between the network structure and the dynamics of this
MJP. Our aim here is to address such questions within the framework of
transition path theory (TPT), originally introduced in~\cite{tpt0}
(see also \cite{tpt1,tpt2} for reviews) and already used
in~\cite{dtpt} in the context of networks and MJPs -- the present work
can be viewed as a continuation of this last paper. In a nutshell, the
basic idea in TPT is to single out two specific sets of nodes and
analyze the statistical properties of the reactive trajectories by
which transitions between these sets occur -- if the sets are chosen
appropriately, this permits to extract the most salient features of
the dynamics on the network and relate them to its topology. This is
like probing an electrical network by wiring it at different locations
and analyzing how the current flow from the nodes wired positively to
those wired negatively~\cite{doysnel,berkon,denhol}.

TPT is also related to the potential-theoretic approach to
metastability championed by Bovier and
collaborators~\cite{bovier0,bovier1,bovier2,bovier3}, albeit the
emphases of both approaches are different. The potential-theoretic
approach has been introduced as a theoretical tool to obtain rigorous
bounds on the low-lying eigenvalues that characterize the slowest
relaxation phenomena in MJPs displaying
metastability~\cite{denholl0,f-w,olivieri}. TPT on the other hand
permits to characterize exactly the statistical properties of the
transition pathways on complex networks that are not necessarily
metastable, or such that the low-lying part of their spectrum is too
complicated to be estimated analytically. Importantly TPT can also be
used as a computational tool in such situations. By being able to
analyze the flow of transitions between specific parts of the network,
for example by generating numerically reactive trajectories by which
these transitions occur, or even no-detour transitions paths, and
analyzing their statistical properties, TPT can provide invaluable
information about the network and the dynamics it supports.

\begin{figure}[t]
\centerline{(a){\includegraphics[height=50mm]{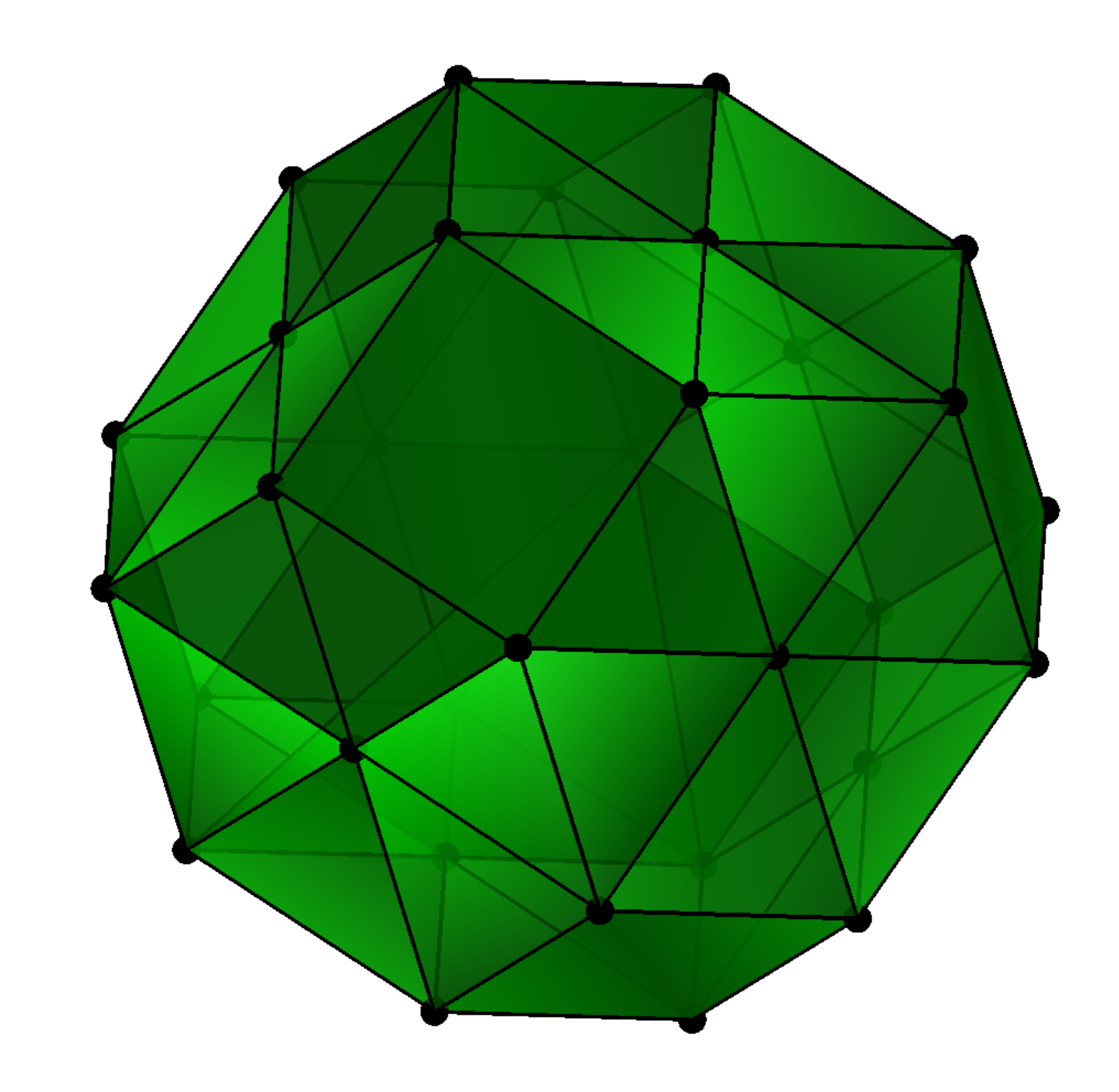}}\hspace{1cm}
(b){\includegraphics[height=50mm]{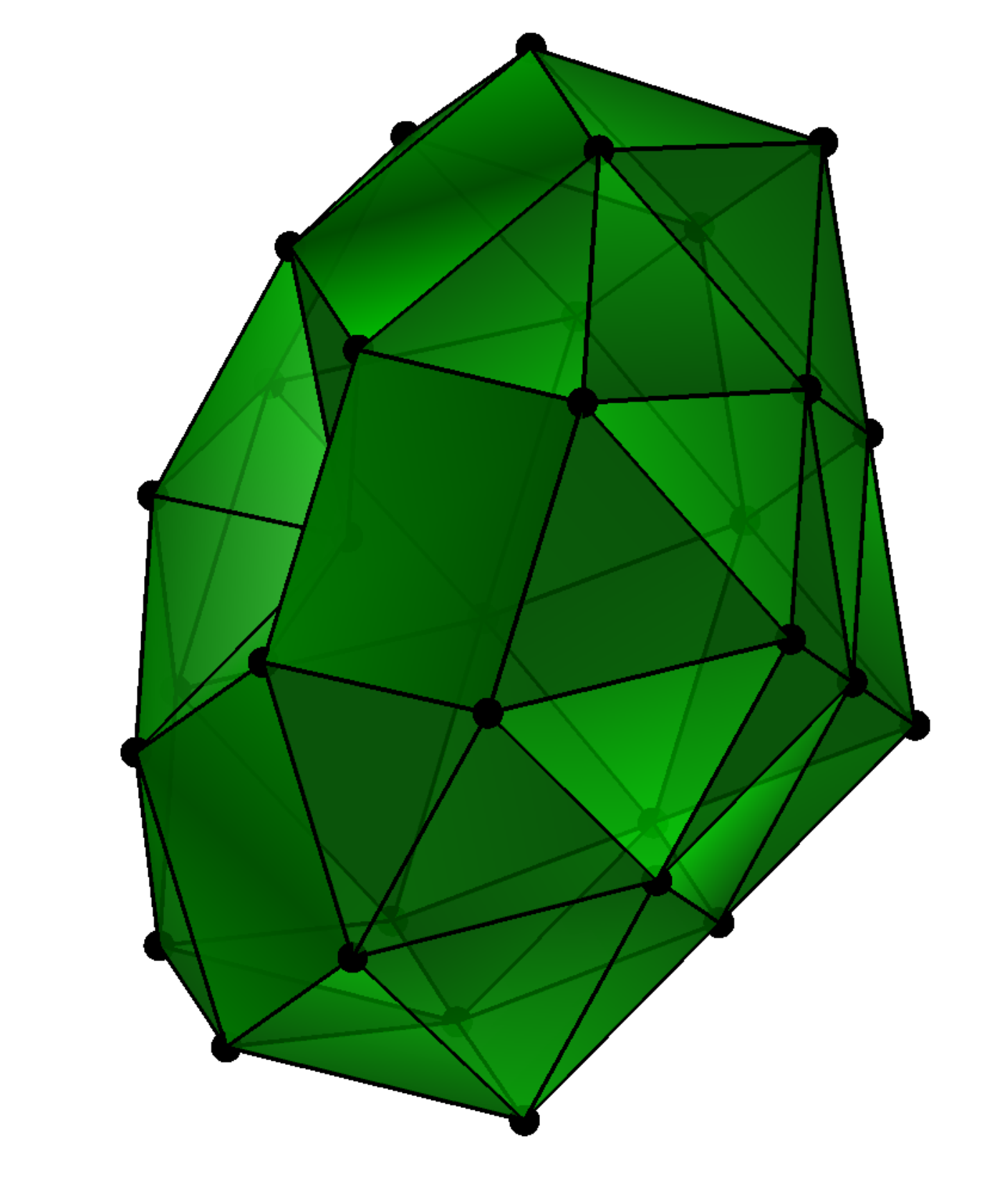}}}
\caption{\label{fcc_ico} The two lowest minima of the potential energy
  of the LJ$_{38}$.  (a): The face-centered cubic truncated octahedron
  with the point group $O_h$ is the lowest minimum.  (b): The
  icosahedral structure with the $C_{5v}$ point group is the second
  lowest minimum.  Throughout this paper we refer to them as FCC and
  ICO, respectively.}
\end{figure}

To make this last point and illustrate the usefulness of the tools
developed in this paper, we will apply them to analyze the network
developed by David Wales and collaborators to model the dynamics of
Lennard-Jones clusters with 38 atoms
($\lj38$)~\cite{wales38,wales-doye}. $\lj38$ is a prototypical example
illustrating how the complexity of a system's energy landscape (and
its associated network) affects its dynamical properties, a feature
that is also observed in other complex phenomena such as protein
folding or glassy dynamics. $\lj38$ has a double-funnel landscape: its
global minimum, a face-centered-cubic truncated octahedron, lies at
the bottom of one funnel, whereas its second lowest minimum, an
incomplete Mackay icosahedron, lies at the bottom of the other (see
Fig.~\ref{fcc_ico}). The deeper octahedral funnel is also narrower,
and believed to be mostly inaccessible from the liquid state. Thus,
when $\lj38$ self-assembles by crystallization, it does so by reaching
the bottom of the shallow but broader isocahedral funnel, and an
interesting question is how does $\lj38$ manage to subsequently find
its ground state structure by travelling from the shallow funnel to
the deep one?  This question of rearrangement is the one that we will
address below. It is made complicated by the ruggedness of the energy
landscape of $\lj38$, which has an enormous number of local minima
separated by a hierarchy of barriers of different heights. 

The remainder of this paper is organized as follows. In
Sec.~\ref{sec:TPT} we summarize the main outputs of TPT. In
Sec.~\ref{sec:current} we introduce sampling tools based on the
theory. In Sec.~\ref{sec:metastable} we discuss the case of metastable
networks, and establish connections between TPT and the potential
theoretic approach to metastability as well as large deviation theory
that arise in these situations. In Sec.~\ref{sec:lj38} we apply the
tools introduced earlier to analyze the rearrangement of the $\lj38$
network. Finally, some concluding remarks are given in
Sec.~\ref{sec:lj38}.

\section{Transition Path Theory}
\label{sec:TPT}

TPT for networks and Markov jump processes (MJPs) is discussed in
detail in~\cite{dtpt} (see also~\cite{tpt_szabo,tpt2}). Here we give a
brief summary of the theory, then discuss algorithms based on it that
can be used to characterize the flows on the network. We also comment
on the connections between TPT and spectral approaches to network
analysis, Bovier's potential theoretic approach to metastability in
MJPs, and large deviation theory.

\subsection{Basic Set-up}
\label{sec:setup}

We will consider MJPs on a countable state-space $S$ with
infinitesimal generator $L =(L_{i,j})_{i,j\in S}$:
\begin{equation}
  \label{ldef}
  \begin{cases}
    L_{i,j}\ge 0,\qquad &\forall i,j\in S,\ i\not=j ,\\
    \sum_{j\in S} L_{i,j} = 0, &\forall i\in S
  \end{cases}
\end{equation}
where $L_{i,j} \Delta t+ o(\Delta t)$ for $i\not=j$ denotes the
probability that the process jumps from state $i$ to state $j$ in the
infinitesimal time interval $[t,t+\Delta t]$. Any such MJP is
equivalent to a network which we denote by $G\{L\} \equiv G(S,E)$: the
set $S$ of states of the MJP is the set of nodes in the network, and
$E$ is the set of edges, i.e. the set of ordered pairs $(i,j)$ with
$i\not = j$ such that $L_{i,j}\not =0$. Conversely, any network with
positive weighted edges is equivalent to an MJP by interpreting the
weights on these edges as off-diagonal entries of the MJP generator.

We assume that the generator is irreducible and that the MJP is
ergodic with respect to the equilibrium probability distribution
$\pi=(\pi_i)_{i\in S}$ satisfying
\begin{equation}
  \label{epi}
  \sum_{i\in S} \pi_i L_{i,j}=0 \quad \forall j \in S, \qquad
  \sum_{i\in S} \pi_i = 1.
\end{equation}
For simplicity, we also assume that the MJP is time-reversible,
i.e. that the detailed balance property holds
\begin{equation}
  \label{detbal}
  \pi_i L_{i,j}=\pi_j L_{j,i}\qquad \forall i,j \in S
\end{equation}
%
We denote by $X(t)$ the instantaneous position of the MJP and
following standard conventions we assume that the function $X(\cdot)$
is right-continuous with left limits (\textit{c\`adl\`ag}).

\subsection{Reactive Trajectories and their Statistical Properties}
\label{sec:react}

TPT is a framework to understand the mechanism by which transitions
from any subset $A\subset S$ to any disjoint subset $B\subset S$ occur
in the MJP. Specifically, TPT analyzes the statistical properties of
the \textit{reactive trajectories} by which these transitions occur:
if $\{X(t)\}_{t\in \RR}$ denotes an infinitely long equilibrium
trajectory of the MJP, the reactive trajectories associated with it
are the successive pieces of $\{X(t)\}_{t\in \RR}$ during which it has
last left $A$ and is on its way to $B$ next. TPT gives explicit
expressions for the probability distribution of the reactive
trajectories, their probability current, their rate of occurrence, etc.

Besides the equilibrium probability distribution $\pi$ and the
generator $L$, the expressions for these quantities involve the
committor $q=(q_i)_{i\in S}$, defined as the probability that the
process starting at a state $i\in S$ will first reach~$B$ rather
than~$A$:
\begin{equation}
  q_i:=\mathbb{P}(\tau_B(i)< \tau_A(i)),
  \label{qdef}
\end{equation}
where $\tau_C(i)$ denotes the first
hitting time of set~$C$ starting from $i$:
\begin{equation}
  \label{eq:hitt}
  \tau_C(i)=\inf\{t\ge0~|~X(0)=i,~X(t)\in C\}
\end{equation}
The committor is also known as equilibrium potential of the capacitor
$(B,A)$, and is denoted by $h_{B,A}$ in the collection of works of
Bovier \textit{et al.}  (see e.g. \cite{bovier1, bovier2,
  bovier3}). It satisfies
\begin{equation}
  \label{lq}
  \begin{cases}
    \sum_{j\in S} L_{i,j}q_j=0,\qquad &\text{if\ $i\in S\backslash(A\cup B)$},\\
    q_i=0, &\text{if\ $i\in A$}\\
    q_i=1, &\text{if\ $i\in B$}
  \end{cases}
\end{equation}
and it can be used to estimate various statistical descriptors of the
reactive trajectories. For example, the equilibrium probability to
find the process in state $i$ and that it be reactive -- which is called
the \textit{probability distribution of reactive trajectories} --
is given by
\begin{equation}
  \label{eq:precat}
  \pi_i^R = q_i (1-q_i) \pi_i
\end{equation}
Indeed, the equilibrium probability to find the trajectory in~$i$ is
$\pi_i$, and the probability that it is reactive, is the product
between $q_i$, which gives the probability that it will reach~$B$
rather than~$A$ next, and $1-q_i$, which by time-reversibility gives
the probability that it came from~$A$ rather than~$B$ last. Note that
$\pi_i^R$ is only non-zero if $i \not\in A\cup B$. Note also that this
distribution is not normalized to one: the quantity
\begin{equation}
  \label{eq:react}
  \rho_R = \sum_{i\in S} \pi_i^R = \sum_{i\in S} q_i (1-q_i) \pi_i
\end{equation}
gives the probability that the trajectory be reactive (i.e. the
proportion of time it spends traveling from $A$ to $B$ at
equilibrium), and the probability to find the trajectory at state $i$
at equilibrium conditional on it being reactive is $\pi_i^R/\rho_R$.

Similarly, we can calculate the average number of transitions per unit
time that the reactive trajectories make from state~$i$ to
state~$j\not=i$:
\begin{equation}
  \label{dpc}
  f_{i,j}^{R}=\begin{cases} (1-q_i)\pi_i L_{i,j}q_j\qquad 
    &\text{if \ $i\neq j$},\\
    0&\text{otherwise}.
\end{cases}
\end{equation}
%
The additional factor $(1-q_i)q_j$ beside the usual $\pi_i L_{i,j}$
accounts for the requirement that, in order to be reactive, the
trajectory must have reached~$i$ coming from~$A$ last and it must
reach~$B$ next after leaving~$j$.  By antisymmetrizing~$f^R_{i,j}$ we
obtain the \textit{probability current of reactive
  trajectories}\footnote{%
  Note that in~\cite{dtpt}, \eqref{dpc} was referred to as the current
  of reactive trajectories and~\eqref{eq:effcurrent} as the effective
  current of reactive trajectories: the terminology used here is more
  consistent with standard conventions in which a current should be
  antisymmetric in its indices.}:
\begin{equation}
  \label{eq:effcurrent}
  F^R_{i,j}=f_{i,j}^{R}-f^{R}_{j,i} = \pi_i L_{i,j} (q_j - q_i).
\end{equation}
This current is key to understand the mechanism of the reaction as it
permits to locate the productive channels by which this reaction
occurs -- in contrast, both~\eqref{eq:precat} and~\eqref{dpc} indicate
where the reactive trajectories go, but these locations may include
many dynamical traps and/or deadends that these trajectories visit but
do not contribute to their current towards~$B$. We will elaborate on
these points in Sec.~\ref{sec:current}. The
current~\eqref{eq:effcurrent} also permits to calculate the average
number of transitions per unit time as the total current out of~$A$ or
into~$B$:
\begin{equation}
  \label{nuab}
  \nu_{R}=\sum_{i\in A,j\in S}F^{R}_{i,j}=\sum_{i\in S,j\in B}F^{R}_{i,j}.
\end{equation}
This quantity is referred to as the \textit{reaction rate} and it can
also be expressed as
\begin{equation}
  \label{nuabsym}
  \nu_{R}=\tfrac{1}{2}\sum_{i,j\in S}\pi_iL_{i,j}(q_j-q_i)^2.
\end{equation}
\eqref{nuabsym} follows from the detailed balance
condition~\eqref{detbal} and the conservation of the current (Theorem
2.13 in \cite{dtpt}): $\sum_{j\in S} F^R_{i,j}=0$ for all $i\in
S$. The reaction rate~$\nu_R$ should not be confused with the rates
$k_{A,B}$ and $k_{B,A}$ defined respectively as the inverse of the average time it
takes the trajectory to go back to $B$ after hitting $A$ or back to $A$
after hitting $B$. These rates are given by
\begin{equation}
\label{rel}
k_{A,B}=\nu_{R}/\rho_A, \qquad k_{B,A}= \nu_R/\rho_B,
\end{equation}
where
\begin{equation}
  \label{eq:2}
  \rho_A=\sum_{i\in S}\pi_i(1-q_i),\qquad\rho_B=\sum_{i\in S}\pi_iq_i
  \qquad (\rho_A+\rho_B=1)
\end{equation}
are the proportions of time such that the trajectory last hit~$A$
or~$B$, respectively.

\section{Sampling and Other Analysis Tools Based on TPT}
\label{sec:current}

In this section we show how the outputs of TPT can be used to
understand the mechanism of the transitions from $A$ to $B$. If we
want to know where these trajectories go, this can be done by
analyzing \eqref{eq:precat} and~\eqref{dpc}. Some of the locations
visited by the reactive trajectories may be deadends, however, in the
sense that not much current goes through them. In order to determine
the productive paths (in term of probability current) taken by the
reactive trajectories, we need to analyze the
current~\eqref{eq:effcurrent}.

Some tools to perform this analysis were already introduced
in~\cite{dtpt}. For example, it was shown how to identify a dominant
representative path, in the sense that this path maximizes the current
it carries. 
While such a path can be informative about the mechanism of the
reaction, it can also be misleading in situations where the
probability current of reactive trajectories is supported on many
paths which carry little current individually -- in other words, in
situations where the reaction channel is spread out. Here we introduce
tools that are appropriate in these situations as well, since we
expect them to be quite generic in complex networks. Specifically, we
provide ways to generate directly reactive trajectories that flow
from~$A$ to~$B$ without even returning to~$A$, or even trajectories
that only take productive steps towards $B$. The statistical
analysis of these trajectories then provides ways to analyze the flows
in the network, which we also discuss.

The following technical assumptions will be used below to simplify
the discussion:
\begin{itemize}
\item[(A)] {\it $L_{i,j}=0$ if $i\in A$ and $j\in B$, i.e. the MJP cannot jump
  directly from $A$ to $B$ -- with this condition, every reactive
  trajectory visits at least one state outside of $A\cup B$.}
\item[(B)] {\it $q_i\not= 0 $ and $q_i\not= 1$ if $i\not \in A\cup
    B$.}
\item[(C)] {\it $q_i\not=q_j$ if $i\not = j$ and $i,j\not \in A\cup B$.}
\end{itemize}
It is straightforward to generalize the statements in
Propositions~\ref{th:tpp} and~\ref{th:lftpp} below to situations where
these assumptions do not hold, as indicated in the proofs, but it
makes them slightly more involved.

\subsection{Transition Path Processes With or Without Detours}
\label{sec:tpp}

Our first result is a proposition that indicates how to generate
reactive trajectories directly. The main idea is to lump onto an
artificial state~$s$ all the pieces of the trajectory in the original
MJP during which it is not reactive.  We call the process obtained
this way the transition path process, following the terminology
introduced in~\cite{lunolen}, where a similar construction was made in
the context of diffusions:

\begin{proposition}[Transition Path Process]
  \label{th:tpp}
  Suppose that assumptions~(A) and (B) hold, let $S_R = S\setminus
  (A\cup B)$, and consider the process on the state-space $\tilde S =
  S_R \cup \{s\}$ defined by the generator with off-diagonal entries
  given by
\begin{equation}
  \label{eq:3}
  \begin{cases}
    \tilde L_{i,j} = L_{i,j} q_j/q_i, \qquad &i,j \not\in A\cup
    B,\ i\not=j, \\
    \tilde L_{i,s} = \sum_{j\in B} L_{i,j}/q_i, & i \not\in A\cup B,\\
    \tilde L_{s,j} = \sum_{i\in A} \pi_i L_{i,j} q_j/(1-\rho_R), 
    & j \not\in A\cup B
  \end{cases}
\end{equation}
where $\rho_R = \sum_{i\in S} q_i (1-q_i) \pi_i$ is the probability
that the trajectory is reactive (see Eq.~\eqref{eq:react}). Then this
process has the same law as the one obtained from the original MJP
by mapping every non-reactive piece of its trajectory onto state~$s$.
In particular, on $S_R$ the invariant probability distribution of the
transition path process coincides with the probability distribution of
the reactive trajectories given in~\eqref{eq:precat}, and the average
number of transition per unit time that the transition path process
makes between states in $S_R$ is given by~\eqref{dpc} and the
associated current by~\eqref{eq:effcurrent}.
\end{proposition}

The proof of this proposition is given at the end of this
section. Note that we can supplement the transition path process with
the information that when it jumps to $j\in S_R$ from $s$, it comes
from state $i\in A$ with probability
\begin{equation}
  \label{eq:6}
  p_{i,j}^{A,S_R} = \frac{\pi_i L_{i,j} q_j}{\sum_{k\in A} \pi_k L_{k,j}
      q_j} = \frac{\pi_i L_{i,j}}{\sum_{k\in A} \pi_k L_{k,j}},
\end{equation} 
and when it jumps to~$s$ from $i\in
S_R$, it reaches state $j\in B$ with probability
\begin{equation}
  \label{eq:4}
  p^{S_R,B}_{i,j} = \frac{L_{i,j}/q_i}{\sum_{k\in B} L_{i,k}/q_i} 
  = \frac{L_{i,j}}{\sum_{k\in B} L_{i,k}}
\end{equation}
With this information added, the invariant probability current of the
transition-path process is the same as the one
in~\eqref{eq:effcurrent} of the reactive trajectories even if we
include edges that come out of~$A$ or into~$B$.

By construction, in the transition path process (like in the reactive
trajectories it represents), the trajectories go from $A$ to $B$
directly, without ever returning to $A$ in between -- in the
transition path process, these returns arise through visits to
state~$s$. In contrast, if we were to simply turn $A$ into a source
and $B$ into a sink, the process one would obtain could take many
steps to travel from $A$ to $B$ because it could revisit $A$ often
before making an actual transition -- this problem is especially acute
if $A$ and $B$ are metastable states since, by definition, $A$ is then
revisited often before a transition to $B$ occurs (more on
metastability in Sec.~\ref{sec:metastable}). In such situations, the
reactive trajectories are much shorter since by construction they only
contain this last transitioning piece. It should be stressed, however,
that the reactive trajectories could still take many steps to travel
from~$A$ to $B$ and be complicated themselves. For example if the
transition mechanism involves dynamical traps or deadends along the
way, the reactive trajectories will wander a long time in the region
between~$A$ and~$B$ before finally making their way to~$B$.

In such situations, it is convenient to construct a process that
carries the same probability current as the reactive trajectories,
but makes no detour to go from~$A$ to~$B$. By this we mean the
following: if we look at the way the committor function varies along a
reactive trajectory, it will start at 0 in~$A$ and go to 1 in~$B$,
but it will not necessarily increase monotonically between these
values along the way. Let us call the pieces of the reactive
trajectories along which the committor increases the productive
pieces, in the sense that they are the ones that bring these
trajectories closer to the product~$B$, whereas they make a detour along
any other piece. Imagine patching together these productive pieces in
such a way that the resulting process is Markov and carries the same
probability current as the reactive trajectories. It turns out that
there is a precise way to do so, and this defines what we call the
no-detour transition path process:

\begin{proposition}[No-Detour Transition Path Process]
  \label{th:lftpp}
  Suppose that assumptions (A), (B), and~(C) hold, let $S_R =
  S\setminus (A\cup B)$ and consider the process on the state-space
  $\tilde S = S_R \cup \{s\}$ defined by the generator with
  off-diagonal entries
\begin{equation}
  \label{eq:3b}
  \begin{cases}
    \hat L_{i,j} = (\rho_{S_R}/\rho_R) L_{i,j} (q_j-q_i)_+, \qquad &i,j
    \not\in A\cup
    B,\ i\not=j, \\
    \hat L_{i,s} = (\rho_{S_R}/\rho_R)\sum_{j\in B} L_{i,j}(1-q_i), & i
    \not\in A\cup
    B,\\
    \hat L_{s,j} = \sum_{i\in A} \pi_i L_{i,j} q_j/(1-\rho_R), & i
    \not\in A\cup B
  \end{cases}
\end{equation}
where $\rho_{S_R} = \sum_{i\in S_R} \pi_i$ and $(q_j-q_i)_+ =
\max\{(q_j-q_i),0\}$. Then this process has the same stationary
current as the transition path process, but the committor function
increases monotonically along each of its paths on $S_R$. In
particular, these paths have no loops.
\end{proposition}

The proof of this proposition is given at the end of this section.
Processes similar to the one in this proposition were introduced in
\cite{bovier_current,denhol}. Note that the equivalent of the
no-detour transition path process for diffusions is somewhat trivial
since the `no-detour' trajectories in this context are simply the
flowlines of the probability current of reactive trajectories, which
are deterministic. Note also that we can again supplement this process
with the information that when it jumps to $j\in S_R$ from $s$, it
comes from state $i\in A$ with probability~\eqref{eq:6}, and when it
jumps to~$s$ from $i\in S_R$, it reaches state $j\in B$ with
probability~\eqref{eq:4}.

\medskip

Propositions~\ref{th:tpp} and \ref{th:lftpp} can be used to generate
reactive trajectories and no-detour reactive trajectories, which can
then be analyzed using a variety of statistical tools to characterize
the mechanism of the reaction. How to do so in practice will be
illustrated on the example of $\lj38$ in
Sec.~\ref{sec:lj38}. Particularly useful is to quantify how these
trajectories go through specific cuts in the network, as we explain in
Sec.~\ref{sec:cutstubes}.

\begin{proof}[Proposition \ref{th:tpp}]
  Under Assumption~(B), the generator $\tilde L$ is irreducible
  because $L$ is. To prove the assertions of the proposition, we will
  verify that the invariant distribution of the transition path
  process is given by
\begin{equation}
  \label{piLtilde} 
  \tilde{\pi}_i=
  \begin{cases}
    q_i(1-q_i)\pi_i,& \text{if}\ \ i\in S_R,\\ 
    (1-\rho_R),& \text{if}\ \ i=s, 
  \end{cases}
\end{equation}
so that the average number of transitions per unit time it makes
between any pair of states, that is, $\tilde f_{i,j} = \tilde \pi_i
\tilde L_{i,j}$, is
\begin{equation}
  \label{ecLtilde}
  \tilde{f}_{i,j} = 
  \begin{cases}
    \pi_i(1-q_i)L_{ij}q_j,& \text{if}\ \  i,j\in S_R,\\
    \sum_{k\in A} \pi_k L_{k,j}q_j,& \text{if}\ \ i=s,~j\in S_R\\
    \pi_i(1-q_i)\sum_{k\in B} L_{i,k},& \text{if}\ \ i\in S_R,~j=s.
  \end{cases}
\end{equation}

To show that \eqref{piLtilde} is the invariant distribution of the
transition path process, we consider two cases: $j\in S_R$ and $j=s$.
For $j\in S_R$ we have
\begin{align*}
  \sum_{i\in S_R \cup\{s\}} \tilde \pi_i \tilde L_{i,j} &=
  \sum_{\substack{i\in S_R \\ i\neq
      j}}\tilde{\pi}_i\tilde{L}_{i,j}+\tilde{\pi}_s\tilde{L}_{s,j}-
  \tilde{\pi}_j\Bigg(\sum_{\substack{i\in S_R\\ i\neq j}}
  \tilde{L}_{j,i}+\tilde{L}_{j,s}\Bigg)\\
  & = \sum_{\substack{i\in S_R\\ i\neq j}
  }\pi_iq_i(1-q_i)L_{i,j}\frac{q_j}{q_i} +   (1-\rho_R)\sum_{i\in A}\frac{\pi_i L_{i,j}q_j}{1-\rho_R}\\
  & \quad - \pi_jq_j(1-q_j)\Bigg(\sum_{\substack{i\in S_R\\ i\neq j}
  }L_{j,i}\frac{q_i}{q_j}+\sum_{i\in
    B}L_{j,i}\frac{1}{q_j}\Bigg)\\
  &=\sum_{\substack{i\in S_R\\ i\neq j}}\pi_i(1-q_i)L_{i,j}q_j
  + \sum_{i\in A}\pi_i L_{i,j}q_j\\
  & \quad - \pi_j(1-q_j)\Bigg(\sum_{\substack{i\in S_R\\ i\neq j}}
  L_{j,i}q_i +\sum_{i\in B}L_{j,i}\Bigg)
\end{align*}
Using the detailed balance condition, $\pi_iL_{i,j}=\pi_jL_{j,i}$, a
few terms cancel out and we are left with
\begin{align*}
  \sum_{i\in S_R \cup\{s\}} \tilde \pi_i \tilde L_{i,j} &=
  \pi_j\sum_{\substack{i\in S_R\\ i\neq j}}L_{j,i}(q_j -q_i) +
  \pi_j\sum_{i\in A} L_{j,i}q_j\\
  &\quad-\pi_j(1-q_j)\sum_{i\in B}L_{j,i}\\
  &= -\pi_j\Bigg(\sum_{\substack{i\in S_R\cup B\\ i\neq j}}
  L_{j,i}q_i-q_j\sum_{\substack{i\in S\\ i\neq j}}L_{j,i}\Bigg)= -
  \pi_j \sum_{i\in S} L_{j,i} q_i=0
\end{align*}
where we used $q_i=0$ if $i\in A$ and $q_i=1 $ if $i\in B$, and the
last equality follows from the definition of the committor. 

For $j=s$ we have
\begin{align*}
  \sum_{i\in S_R\cup \{s\}}\tilde{\pi}_i\tilde{L}_{i,s} &= \sum_{i\in
    S_R}\tilde{\pi}_i\tilde{L}_{i,s}-\tilde{\pi}_s
  \sum_{i\in S_R}\tilde{L}_{s,i}\\
  &= \sum_{i\in S_R}q_i(1-q_i)\pi_i\sum_{k\in B}L_{i,k}\frac{1}{q_i}-(1-\rho_R)\sum_{i\in S_R}\frac{\pi_kL_{k,i}q_i}{1-\rho_R}\\
  &= \sum_{i\in S_R}\sum_{k\in B}L_{i,k}\pi_i(1-q_i)-\sum_{k\in
    A}\sum_{i\in S_R}L_{k,i}\pi_kq_i\\
  &= \nu_R-\nu_R=0,
\end{align*} 
which terminates the proof. Note that if Assumption (A) does not hold,
then we also need to account for the direct jumps from $A$ to $B$ in
the original MJP as additional visits into state~$s$. If Assumption
(B) does not hold, we can fatten the states $A$ and $B$ to include all
the nodes $i$ such that $q_i=0$ and $q_i=1$, respectively.  With this
modification, the proposition is valid.
\end{proof}


\begin{proof}[Proposition \ref{th:lftpp}] The fact that the process
  has no loops follows directly from the form of its generator -- in
  particular the network defined by $\hat L$, $G\{\hat L\}$, has no
  loops except for the ones through~$s$.  The proof of the rest of the
  statement is similar to that of Proposition~\ref{th:tpp}: Under
  Assumptions~(B) and (C), the generator $\hat L$ is irreducible
  because $L$ is and we will show that the invariant distribution in
  the network with the generator $\hat{L}=(L_{i,j})_{i,j\in S_R \cup
    \{s\}}$ in~\eqref{eq:3b} is equal to
\begin{equation}
  \label{piLhat}
  \hat{\pi}_i=
  \begin{cases}
    \pi_i \rho_R/\rho_{S_R},&\text{if}\ \ i\in S_R,\\ 
    1-\rho_R,&\text{if}\ \ i=s,
  \end{cases}
\end{equation}
so that the average number of transitions per unit time it makes
between any pair of states, that is, $\hat f_{i,j}= \hat \pi_i \hat
L_{i,j}$, is
\begin{equation}
  \label{ecLhat}
  \hat{f}_{i,j}=
  \begin{cases}
    \pi_iL_{i,j}(q_j-q_i)_+,&\text{if}\ \ i,j\in S_R,\\
    \sum_{k\in A}\pi_kL_{k,j}q_j,&\text{if}\ \ i=s,~j\in S_R,\\
    \pi_i(1-q_i)\sum_{k\in B}L_{i,k},&\text{if}\ \ i\in S_R,~j=s.
  \end{cases}
\end{equation}
This will imply that the transition path process and the no-detour
transition path process have the same stationary current, as claimed
in the proposition.

To show that~\eqref{piLhat} is the invariant distribution, we consider
again two cases: $j\in S_R$ and $j=s$.  If $j\in S_R$ we have
\begin{align*}
  \sum_{i\in S_R\cup \{s\}} \hat \pi_i \hat L_{i,j} &=
  \sum_{\substack{i\in S_R\\ i\neq j}
  }\hat{\pi}_i\hat{L}_{i,j}+\hat{\pi}_s\hat{L}_{s,j}-
  \hat{\pi}_j\Bigg(\sum_{\substack{i\in S_R\\ i\neq j}}
  \hat{L}_{j,i}+\hat{L}_{j,s}\Bigg)\\
  &= \sum_{\substack{i\in S_R\\ i\neq
      j}}\frac{\pi_i\rho_R}{\rho_{S_R}}
  L_{i,j}(q_j-q_i)_{+}\frac{\rho_{S_R}}{\rho_{R}}
  +(1-\rho_R)\sum_{i\in A}\frac{\pi_i L_{i,j}q_j}{1-\rho_R}\\
  & \quad - \frac{\pi_j\rho_R}{\rho_{S_R}}\Bigg(\sum_{\substack{i\in S_R\\ i\neq
      j}}L_{j,i}(q_i-q_j)_{+}\frac{\rho_{S_R}}{\rho_{R}}
  +\sum_{i\in B}L_{j,i}(1-q_j) \frac{\rho_{S_R}}{\rho_{R}}\Bigg)
\end{align*}
Using the detailed balance condition, $\pi_iL_{i,j}=\pi_jL_{j,i}$, and
the fact that
$$
(q_j-q_i)_{+}-(q_i-q_j)_{+}=q_j-q_i
$$
we obtain
\begin{align*}
  \sum_{i\in S_R\cup \{s\}} \hat \pi_i \hat L_{i,j} &=
  \pi_j\sum_{\substack{i \in S_R\\ i\neq j}}L_{j,i}(q_j-q_i)
  +\pi_j\sum_{i\in A} L_{j,i}q_j
  -\pi_j\sum_{i\in B}L_{j,i}(1-q_j)\\
  &= -\pi_j\Bigg( \sum_{\substack{i\in S_R\cup B\\ i\neq
      j}}L_{j,i}q_i-q_j\sum_{\substack{i \in S\\
      i\neq j}}L_{j,i}\Bigg)=-\pi_j \sum_{j\in S} L_{j,i} q_i = 0
\end{align*}
where we used $q_i=0$ if $i\in A$ and $q_i=1 $ if $i\in B$, and the
last equality follows from the definition of the committor.

For $j=s$ we have
\begin{align*}
  \sum_{i\in S_R\cup \{s\}} \hat \pi_i \hat L_{i,s} &=
  \sum_{i\in S_R}\hat{\pi}_i\hat{L}_{i,s}-\hat{\pi}_s\sum_{i\in S_R}\hat{L}_{s,i}\\
  &= \sum_{i\in S_R}\frac{\pi_i\rho_R}{\rho_{S_R}}
  \sum_{k\in B}L_{i,k}(1-q_i)\frac{\rho_{S_R}}{\rho_R}
  -(1-\rho_R)\sum_{i\in S_R}\frac{\pi_kL_{k,i}q_i}{1-\rho_R}\\
  & =\sum_{i\in S_R}\sum_{k\in B}L_{i,k}\pi_i(1-q_i)-\sum_{k\in
    A}\sum_{i\in S_R}L_{k,i}\pi_kq_i\\
  & = \nu_R-\nu_R=0.
\end{align*}
which ends the proof. To remove Assumption (A), we need to account for
the direct jumps from $A$ to $B$ in the original MJP as additional
visits into state~$s$. To remove Assumption (B), we can fatten the
states $A$ and $B$ to include into them all the nodes~$i$ such that
$q_i=0$ and $q_i=1$, respectively. And to remove Assumption (C), we
can restrict the statement of the proposition to the unique ergodic
component of the chain with generator~\eqref{eq:3b} composed of all
the states in~$S_R$ that can be reached starting from $A$. 

\end{proof}

\subsection{Isocommittor cuts and transition channels}
\label{sec:cutstubes}

\begin{figure}[htbp]
  \centerline{\includegraphics[width=0.5\textwidth]{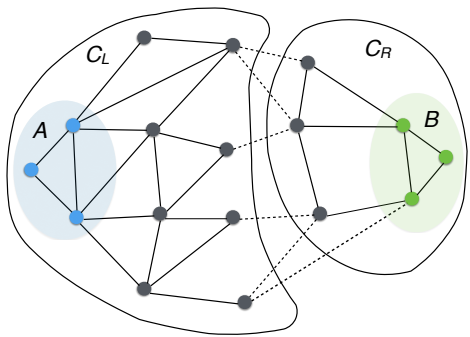}}
  \caption{%
    Illustration for the concept of an $A$-$B$-cut between the sets
    $A$ and $B$ whose nodes are shown in blue and green respectively . The
    edges of the cut-set are shown with dashed lines.}
  \label{fig:cut}
\end{figure}

Recall that a cut in a network $G(S,E)$ is a partition of the nodes
in~$S$ into two disjoint subsets that are joint by at least one edge
in~$E$. The set of edges whose endpoints are in different subsets of
the partition is referred to as the cut-set. Here we will focus on
$A$-$B$-cuts that are such that $A$ and $B$ are on different
sides of the cut-set.  Any $A$-$B$-cut leads to the
decomposition $S=C_L\cup C_R$ such that $C_L\supseteq A$ and
$C_R\supseteq B$ (see Fig. \ref{fig:cut}).

We can use cuts to characterize the width of the transition tube
carrying the current of reactive trajectories. A specific set of cuts
is convenient for this purpose, namely the family of isocommittor cuts
which are such that their cut-set $C$ is given by
\begin{equation}
  \label{seq}
  C(q^{\ast})=\{(i,j)~|~q_i\le q^{\ast}, q_j>q^{\ast}\}, \qquad q^{\ast}\in[0,1).
\end{equation}
The isocommittor cuts are the counterparts of the isocommittor
surfaces in the continuous case. These cuts are special because if
$i\in C_L$ and $j\in C_R$, the reactive current between these nodes is
nonnegative, $F_{i,j}^R\ge 0$, which also means that every no-detour
transition path contains exactly one edge belonging to an isocommittor
cut since the committor increases monotonically along these transition
paths.  Therefore, we can sort the edges in the isocommittor cut
$C(q)$ according to the reactive current they carry, in descending
order, and find the minimal number of edges $N(q)$ carrying at least
$p$\% of this current. By doing so for each value of the committor
$0\le q< 1$ and for different values of the percentage
$p\in(0,100)$, one can then analyze the geometry of the transition
channel - how broad is it, how many sub-channels there are, etc. The
result of this procedure will also be illustrated on the example of
$\lj38$ in Sec.~\ref{sec:lj38}.

Finally note that the reaction rate can be expressed as the total
current through any cut (not necessarily an isocommittor cut) as
(compare~\eqref{nuab})
\begin{equation}
  \label{rate2}
  \nu_{R}=\sum_{i\in C_L,j\in C_R} F^R_{i,j}.
\end{equation}
The proof of this statement is elementary and will be omitted.



\section{The Case of Metastable Networks}
\label{sec:metastable}

In this section, we briefly discuss the case of metastable
networks. We start by giving a spectral definition of metastability,
then discuss the connections of our results to the potential theoretic
approach to metastability and to large deviation theory.

\subsection{Spectral Definition of Metastability}
\label{sec:spectral}

Metastable networks and MJPs have been the subject of many studies
(e.g.~\cite{landim1,bovier0,denholl0,olivieri}).  By definition, they
are such that the spectrum of their generator contains one or more
groups of low-lying eigenvalues. Let us assume without loss of
generality that $S = \NN_0$ or $S=\{0,1,\ldots, N\}$ and denote by
$\{(\phi_i^k,\lambda_k)\}_{k\in S}$ the solutions of the eigenvalue
equation
\begin{equation}
  \label{eq:7}
  \sum_{j\in S} L_{i,j} \phi_j = - \lambda \phi_i.
\end{equation}
Then the detailed balance condition~\eqref{detbal} implies that the
eigenvalues are real, non-negative, and can be ordered as $0=\lambda_0 <
\lambda_1 \le \lambda_2 \le \ldots$ There is a low-lying group of
eigenvalues if there exists a $P\in \NN$ and an $\delta \ll1$ such
that
\begin{equation}
  \label{eq:10}
  \lambda_{P-1}/\lambda_P < \delta.
\end{equation}
To see that this condition implies metastability, notice that the
spectral decomposition of the generator,
\begin{equation}
  \label{eq:8}
  L_{i,j} = -\sum_{p\in S} \lambda_p \phi_i^p \phi_j^p \pi_j,
\end{equation}
leads to the following expression for the transition probability
distribution $(e^{t L})_{i,j}$ to find the walker at state~$j$ at
time~$t\ge0$ if it was at~$i$ initially:
\begin{equation}
  \label{eq:9}
  (e^{t L})_{i,j} = \sum_{p\in S} e^{-\lambda_p t} \phi_i^p \phi_j^p \pi_j
\end{equation}
If \eqref{eq:10} holds, it means that on time-scales such that
$\lambda_{P-1} t = O(1)$ in $\delta$, we have $\lambda_P t =
O(\delta^{-1})$, and up to errors that are exponentially small in
$\delta^{-1}$, the sum in~\eqref{eq:9} can effectively be truncated
at $P-1$:
\begin{equation}
  \label{eq:9trunc}
  (e^{t L})_{i,j} = \sum_{p<P} e^{-\lambda_p t} \phi_i^p \phi_j^p
  \pi_j + O(e^{-\delta^{-1}})
\end{equation}
In other words, on these time scales the fast processes described by
the eigenvalues of index $P$ and above have already relaxed to
equilibrium and what remains are the slow processes associated with
the eigenvalues of index $P-1$ and below. This also means that the
dynamics on these time scales can effectively be reduced to a Markov
jump processes on a state space with $P$ states. Note that the
spectral decomposition in~\eqref{eq:9} also leads to a spectral
decomposition for the current: 
\begin{equation}
  \label{eq:11}
  \begin{aligned}
    \frac{d }{dt} (e^{t L})_{i,j} &= \sum_{k\in S} (e^{t L})_{i,k}
    L_{k,j}\\
    & = \sum_{k\not= j } \left( (e^{t L})_{i,k} L_{k,j} - (e^{t
        L})_{i,j} L_{j,k}\right)\\
    & = - \sum_{p\in S} e^{-\lambda_p t} \phi_i^p
    \sum_{k\not= j }  F_{k,j}^p
  \end{aligned}
\end{equation}
where the eigencurrent associated with the pair $(\phi_i^p,\lambda_p)$ is 
\begin{equation}
  \label{eq:12}
  F_{i,j}^p = \pi_i L_{i,j} (\phi_j^p-\phi_i^p)
\end{equation}

\subsection{Potential Theoretic Approach to Metastability}
\label{sec:pot}

The eigencurrent~\eqref{eq:12} should be compared
to~\eqref{eq:effcurrent}: as can be seen, \eqref{eq:12} can be
obtained from \eqref{eq:effcurrent} by substituting the eigenvector
$\phi_i^p$ for the committor~$q_i$.  This suggests that if $p< P$ and
\eqref{eq:12} corresponds to an eigencurrent associated with a slow
process in the low-lying group, then it should be possible to find
sets $A$ and $B$ such that the current of reactive trajectories
$F_{i,j}^R$ between these two sets approximates $F_{i,j}^p$. This is
indeed the case, and this observation is at the heart of the potential
theoretic approach to metastability developed by Bovier and
collaborators~\cite{bovier0,bovier1,bovier2,bovier3}. In a nutshell,
this approach says that, up to shifting and scaling, any low lying
eigenvector $\phi^p_i$ can be approximated by the committor function
for the reaction between two suitably chosen sets $A$ and $B$. 
This observation is useful for analysis because it permits to focus on
a specific eigenfunction/eigenvalue pair by studying the variational
problem that the committor satisfies, that is, by minimizing the
Dirichlet form associated with the generator~$L$:
\begin{equation}
  \label{eq:5}
  \Phi(\tilde q)= 
  \frac12 \sum_{i,j\in S} \pi_i L_{i,j} (\tilde q_j - \tilde q_i)^2
\end{equation}
over all $\tilde q= (\tilde q_i)_{i\in S}$ subject to the boundary
conditions that $\tilde q_i =0$ if $i\in A$ and $\tilde q_i = 1$ if
$i\in B$. The minimizer of~\eqref{eq:5} is the committor function and,
by~\eqref{nuabsym}, its minimum is also the reaction rate~$\nu_R$.

The discussion above makes a (brief) connection between the potential
theoretic approach to metastability and TPT. In fact, TPT gives a way
to reinterpret the various objects used in the potential theoretic
approach in terms of exact statistical descriptors of the reactive
trajectories.  This reinterpretation is interesting because TPT
applies regardless on whether the system is metastable or not. In
other words, all of the formulas given in Secs.~\ref{sec:react}
and~\ref{sec:current} are exact no matter what the sets~$A$ and~$B$
are. This has the advantage that we can use the tools of TPT to
analyze reactions even in situations where \eqref{eq:10} does not
necessarily hold. More generally, our emphasis is different: we are
mainly interested in using TPT to compute numerically the pathways for
a reaction of interest between sets that are known before hand, rather
than estimating analytically the low lying part of the
spectrum. Indeed, while this second goal rapidly becomes out of reach
in practice for complex systems (and typically require to make
specific assumptions about the network like e.g. the ones discussed in
Sec.~\ref{sec:ldt} below), the first one remains achievable in a much
broader class of situations, as will be illustrated in
Sec.~\ref{sec:lj38} on the specific example of $\lj38$.

\subsection{Large Deviation Theory (LDT)}
\label{sec:ldt}

Another question of interest is when does condition~\eqref{eq:10}
applies? One such situation occurs when the state-space is finite, $S
=\{1,2,\ldots, N\}$, and the pairwise rates $L_{i,j}$ are
logarithmically equivalent to $\exp(-U_{i,j}/\epsilon)$ in the limit
as $\epsilon\to0$. The asymptotic properties of the eigenvalues in
such systems, not necessarily with detailed-balance, was first
established by A. Wentzell \cite{wentzell2} using the tools from large
deviation theory (LDT) developed in~\cite{freidlin-cycles} and
summarized in~\cite{f-w} (see also~\cite{olivieri}).

Here we will focus on a sub-case of the one investigated by Wentzell
which is relevant in the context of $\lj38$, namely, when the
generator of the MJP is of the form
%
\begin{equation}
  \label{eq:13}
  L_{i,j} = \frac{\nu_{i,j}}{\nu_i} \exp\left(-\frac1\epsilon(V_{i,j} - V_i)
  \right) 
\end{equation}
where $\nu_{i,j}=\nu_{j,i}>0$, $\nu_i>0$, $V_{i,j} = V_{j,i}>
\max\{V_i,V_j\}$ and $V_i$ are parameters. The generator~\eqref{eq:13}
corresponds to a dynamics on the network where every node~$i\in S$ has
an energy~$V_i$ associated with it, and jumps between adjacent nodes
on the network follow Arrhenius law, with a rate depending
exponentially on the energy barrier $V_{i,j}-V_i$ to hop from~$i$
to~$j$: the information about the network topology is embedded in the
energies by setting $V_{i,j}=+\infty$ if $i$ and $j$ are not adjacent
on the network, i.e if $(i,j)\not \in E$.  The parameter $\epsilon$
plays the role of the temperature, and $\nu_{i,j}/\nu_i$ is a
prefactor which we will assume temperature-independent. The
generator~\eqref{eq:13} satisfies the detailed balance
condition~\eqref{detbal} with respect to be Boltzmann-Gibbs
equilibrium probability distribution
\begin{equation}
  \label{eq:14}
  \pi_i = Z^{-1} \nu_i \exp\left(-\frac1{\epsilon} V_i
  \right), \qquad Z = \sum_{i\in S} \nu_i \exp\left(-\frac1{\epsilon} V_i
  \right)
\end{equation}

\begin{figure}[t]
  \centerline{\includegraphics[width=1\textwidth]{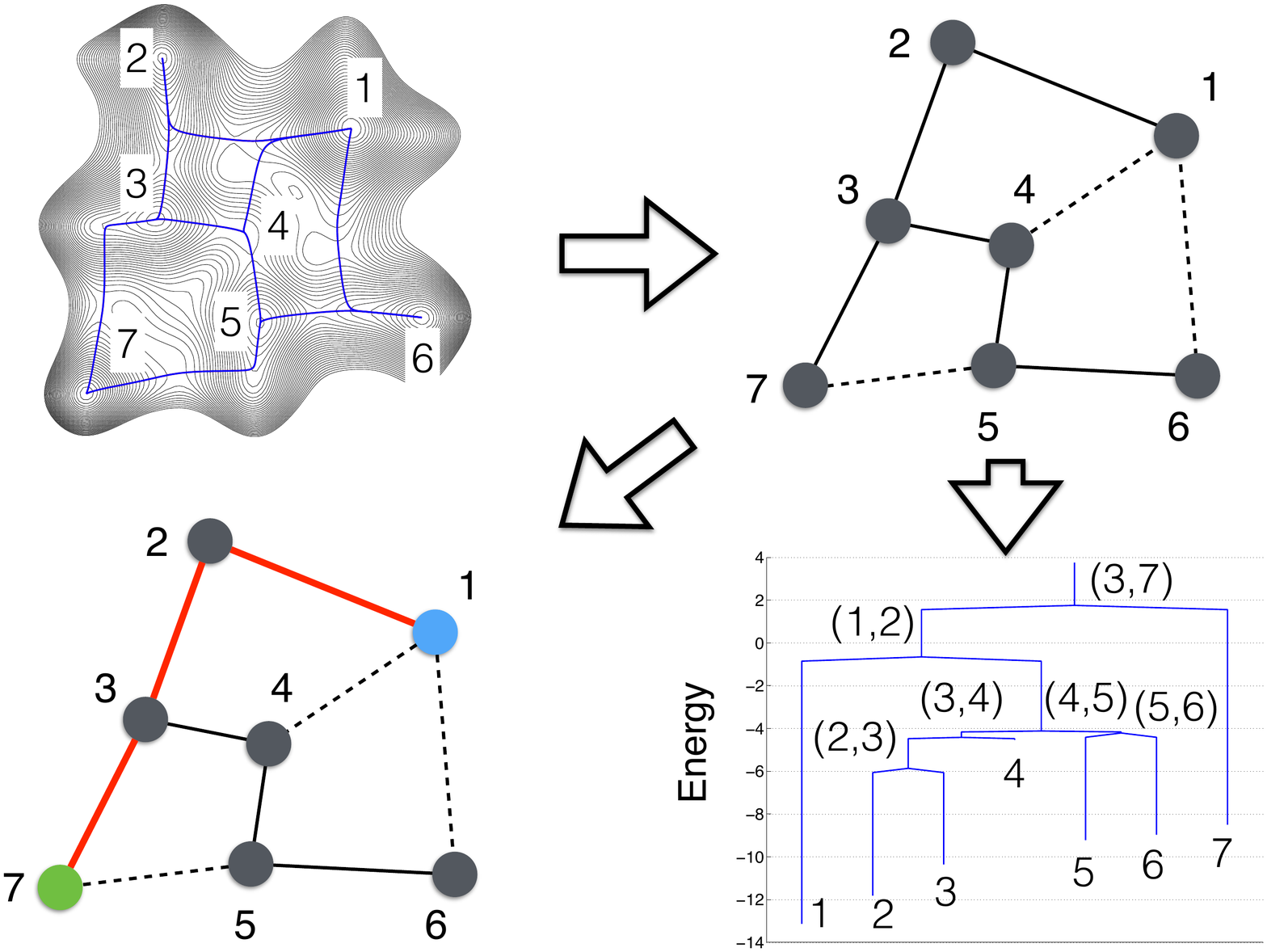}}
  \caption{A continuous 7-well potential (top left) is mapped onto a
    discrete network (top right) by identifying the minimum energy
    paths (MEPs) connecting the local minima of the potential via
    saddle points.  The states $i$, $j$, ... are the indices of these
    minima, there is an edge between any pair $(i,j)$ if there is a
    MEP with a single saddle point along it connecting $i$ and $j$. By
    using the energy of the saddle point $V_{i,j}$ as cost for the
    edge $(i,j)$, one can find the minimal spanning tree of the
    network (solid edges in the top right panel) using e.g. Kruskal
    algorithm, and thereby obtain its disconnectivity graph (bottom
    right). On this disconnectivity graph, the pairs of numbers at the
    branching points indicate which of the nodes in the corrsponding
    bottom parts of the tree connect at that level of energy.  Using
    the Dijkstra-based algorithm proposed in~\cite{cam1} we can also
    calculated the minmax path connecting two states, for example
    between states 1 and 7 (solid red path in the bottom left
    panel). This minmax path is relevant in regimes where LDT
    applies. }
  \label{fig:7well}
\end{figure}

In the set-up above, we can use the temperature~$\epsilon$ as control
parameter, in such a way that~\eqref{eq:10} holds when
$\epsilon\to0$. In that limit, for reasons that will become clear
below, in general there are as many low-lying groups of eigenvalues as
there are states (i.e. $\lambda_p/\lambda_{p+1} \to0$ as
$\epsilon\to0$ for all $p=0,1,\ldots, N-1$), and Wentzell's approach
provides a way to estimate each of these eigenvalues. To see how, it
is convenient to organize the states of the chain on a disconnectivity
graph, that is, a downward facing tree in which each node $i\in S$
lies at the end of a branch at a depth equal to its energy $V_i$, and
branches in the tree are connected at the lowest energy barrier
$V_{j,k}$ that connect all the nodes on one side of the tree to those
on the other side -- a cartoon example is shown in
Fig.~\ref{fig:7well}: because this will be relevant in our analysis of
$\lj38$, in this example we start from a continuous energy landscape
that we convert into a network whose disconnectivity graph is then
obtained from its minimal spanning tree (bottom right, all solid
edges) calculated using e.g. Kruskal's algorithm (see e.g.~\cite{amo}). The
eigenvalues can then be estimated recursively from the disconnectivity
graph as follows: Start by identifying the lowest barrier in the tree,
i.e. the adjacent pair $(i_N,j_N)$ on the tree such that
$V_{i_N,j_N}-V_{i_N}$ is minimum over all $i,j\in S$. The node $i_N$
identifies the well that the system can escape by crossing the barrier
of minimum height, and the largest eigenvalue in the system
corresponds to the inverse of the time scale of this escape, i.e. it
can then be estimated as
\begin{equation}
  \label{eq:15}
  \lambda_{N-1} \asymp \exp\left(-\frac1\epsilon (V_{i_N,j_N}-V_{i_N})\right)
\end{equation}
where the symbol $\asymp$ means that the ratio of the logarithms of
both sides of this equality tends to 1 as $\epsilon\to0$. Now remove
the node $i_N$ and its branch from the tree, and repeat the
construction: that is, in the new tree find the pair
$(i_{N-1},j_{N-1})$ such that $V_{i_{N-1},j_{N-1}}-V_{i_{N-1}}$ is
minimum over all $i,j\in S\setminus \{i_N\}$, to obtain an estimate
for the next largest eigenvalue, $\lambda_{N-2}$.  By
iterating upon this procedure, in $N-1$ steps we can then estimate
$\lambda_{N-k}$ for $k=2,3,\ldots, N-1$ as
\begin{equation}
  \label{eq:15bbb}
  \begin{aligned}
    & \lambda_{N-k} \asymp \exp\left(-\frac1\epsilon
      (V_{i_{k},j_k}-V_{i_k})\right) \qquad \text{where}\\
    & (V_{i_k,j_k}-V_{i_k}) = \min_{i,j \in S
      \setminus\{i_N,i_{N-1},\ldots, i_{N-k+1}\}}(V_{i,j}-V_{i}),
  \end{aligned}
\end{equation}
Intuitively, this procedure corresponds to lumping together the states
that can be be reached on timescales of order $\lambda^{-1}_{k}$ or
below, and analyzing what happens on the next timescale to get
$\lambda_{k-1}\ll \lambda_k$. After $N-1$ steps in the procedure we
end up with a degenerate tree made of a single node lying at the very
bottom of the original tree (and of course we already know that
$\lambda_0=0$).  Note that in the discussion above, we assumed that
the barriers $(V_{i_{k},j_{k}}-V_{i_{k}})$ identified along the way
are all different (that is, strictly increasing with $k$), which is
the generic case and leads to eigenvalues that are all well-separated:
if some of these barriers are equal, it means that some of the
eigenvalues are asymptotically equivalent, and this case can be
treated as well by generalizing the construction above. Note also that
estimates more precise than~\eqref{eq:15} and \eqref{eq:15bbb} can be
obtained using the potential theoretic/TPT approach: in the present
situation, at any stage in the iteration procedure, the states~$i_k$
and~$j_k$ are those that should be set as~$A$ and~$B$, respectively.

\subsubsection{Freidlin's Cycles and MinMax Paths.}
Another interesting construction provided by LDT is the decomposition
of the stochastic network into Freidlin's cycles
\cite{freidlin-cycles,f-w,freidlin-physicad}. For systems satisfying
the detailed balance condition and with a rate matrix as
in~\eqref{eq:13}, the decomposition into cycles simplifies, as was
recently discussed in~\cite{cam1}. Here we summarize this discussion
and refer the interested reader to the original paper for details.

In a nutshell, the decomposition into cycles focuses on which states
are most likely to be reached from a given state: in the  zero
temperature limit, if the system is in state $i$, with probability one
it will reach next  the state connected to $i$ by the smallest
barrier, i.e. 
\begin{equation}
  \label{eq:16}
  j_*(i) = \mathop{\text{arg min}}\limits_{i\in S} V_{i,j} 
\end{equation}
Searching consecutively for the next most likely state defines a
dynamics on the network that generically ends with cycles made of two
states: each of these cycles contain a local minimum of energy on the
disconnectivity graph (that is, a state at the bottom of a group of
branches on the tree), and the state connected to this minimum by the
lowest barrier. These cycles are called 1-cycles by Freidlin. Once we
have identified them, we can remove from the tree the state with
highest energy in each of these 1-cycles, and repeat the construction
iteratively. These gives 2-cycles, 3-cycles, etc. until we again end
up with a tree with only 2 nodes on it. In this construction, we can
also keep the information about the state in the original network by
which any $n$-cycle is exited: with probability 1 as $\epsilon\to0$,
this is the state whose barrier is the lowest to escape all the states
contained in this $n$-cycle.  A corollary of the fact that cycles are
exited in a predictable way is that, between any two nodes in the
network taken as sets $A$ and $B$, there exists a single path on the
network that concentrates all the current of the reactive trajectories
as $\epsilon\to0$. This path has a minmax property: the
maximal barrier separating every pair of states $i$ and $j$ on the
path is minimal among the maximal barriers along all paths in the
network connecting $i$ and $j$ (see Fig.~\ref{fig:7well} for an
illustration).

In~\cite{cam1}, the construction of the hierarchy of Freidlin's cycles
was performed via a sequence of conversions of rate matrices into jump
matrices followed by taking limits $\epsilon\to 0$.  Relying on the
properties of the hierarchy of cycles specific for the systems with
detailed balance, an efficient Dijkstra-based algorithm was also proposed
for computing the minmax path.  Importantly, this algorithm did not
built the whole hierarchy of cycles, but only computed the
sub-hierarchy relevant to the transition process of interest, and did
not require any pre-processing of the stochastic network.

We conclude this section on LDT with a remark. As explained above, the
LDT picture applies in the limit when $\epsilon\to0$, in which case
the hierarchy of different barriers in the disconnectivity graph
corresponds to timescales that become infinitely far apart as
$\epsilon\to0$. While this picture is indeed correct at extremely low
temperature, we do not expect it to remain valid as the temperature is
increased, even if the system does remain strongly metastable
(i.e. such that some low-lying groups of eigenvalue do
persist). Rather, we expect that the transition channel will rapidly
broaden if the networks is large, and that the mechanism of the
reaction wil depart from that predicted by LDT. Our analysis of
$\lj38$ by TPT will indeed confirm this picture.

\section{Application to the Rearrangement of the Lennard-Jones 38
  Cluster}
\label{sec:lj38}

\subsection{Microscopic Model and Thermodynamic Properties}
\label{sec:thermo}

A Lennard-Jones cluster is made of particles (or atoms) interacting
via the Lennard-Jones pairwise potential given by
\begin{equation}
  \label{eq:LJpe}
  V(r) = 
  4a
  \sum_{i<j}\left[\left(\frac{\sigma}{r_{ij} }\right)^{12} 
    - \left(\frac{\sigma}{r_{ij} }\right)^{6}\right].
\end{equation}
Here $r=\{r_j\}_{j=1}^{N}\in \RR^{3N}$ denotes the positions of the
$N$ particles in the cluster, $r_{ij}$ is the distance between
particles $i$ and $j$, and $a>0$ and $\sigma>0$ are parameters
measuring respectively the strength and range of the interactions. At
the most fundamental level, the finite-temperature dynamics of the
cluster can be modeled as a continuous diffusion over the
potential~\eqref{eq:LJpe}. This dynamics is extremely complicated
owing to the multiscale nature of this potential which, when $N$ is
large (e.g. $N=38$ as we will consider below) possesses an enormous
number of local minima separated by a hierarchy of barriers of various
height. A few thermodynamic properties of these clusters are known,
however.

First, it is known that the majority of global potential energy minima
for Lennard-Jones clusters of various sizes involve an icosahedral
packing \cite{wales-doye}.  However, Lennard-Jones clusters with
special numbers of atoms admit a high symmetry configuration based on
a face-centered cubic packing, with a lower energy.  The smallest
cluster with this property contains $N=38$ atoms\cite{wales38,
  wales_landscapes}. The global potential energy minimum of the
$\lj38$ cluster is achieved by a truncated octahedron with the point
group $O_h$ (Fig.~\ref{fcc_ico} (a)), which from now on we will simply
refer to as FCC.  The second lowest minimum is the icosahedral
structure with the $C_{5v}$ point group (Fig. \ref{fcc_ico} (b)),
which we will refer to as ICO.

It is also known that the basin around ICO is much wider than that
around FCC -- these two basins are usually referred to as funnels in
the literature. This has thermodynamic consequences when the
temperature of the system is non-zero.  Indeed, the FCC basin only
remains the preferred basin for $T < T_c$ with $k_B T_c \approx 0.12
a $ (here $k_B $ denotes Boltzmann constant).  At $T=T_c$, the
system undergoes a solid-solid phase transition where the ICO basin
becomes more likely due to its with greater configurational entropy
(see e.g. Fig. 4 in \cite{wales_landscapes}).  Next, at
$k_BT\approx 0.18a$, the outer layer of the cluster melts, while the
core remains solid.  Then the cluster completely melts at $k_B
T\approx 0.35a$ \cite{frantsuzov}.

The difference of widths of the two basins also has dynamical
consequences. Indeed, due to its larger width, the ICO basin is the
one that is most likely to be reached by the system after
crystallization even if $T < T_c $. The question then becomes how does
the system reorganize itself to get out of the dynamical trap around
ICO and in its preferred state around FCC? It is also of interest to
understand how this process is influenced by the temperature, since
the rearrangement pathway is likely to be influenced by it. These are
the type of questions that we will address in this section, as an
illustration of the TPT-based network analysis tools presented
earlier. This study is complementary to those conducted by Wales and
collaborators in the same context~\cite{wales38} using
different tools~\cite{wales0,wales1,wales_landscapes,walesgraph}.

\subsection{Network Representation of the  Lennard-Jones 38 Cluster}
\label{sec:MSM} 

The problem of rearrangement of $\lj38$ has been the object of much
studies in the past 15 years (see
e.g. \cite{wales38,wales-doye,picciani,neirotti,miller}).  An
interesting approach to the problem has been proposed by David Wales
and collaborators, who undertook an ambitious program aiming at
mapping the evolution of $\lj38$ onto a network/MJP and reducing the
analysis of the dynamics of $\lj38$ to the study of this
network. While this mapping is technically hard to perform in practice
and required a lot of inventiveness, it is conceptually quite simple
to understand. If the temperature of the system is small enough, it
will spend a long time near the bottom of the energy well around the
local minima it is currently in before a thermal fluctuation large
enough will manage to push it above an energy barrier separating it
from an adjacent well. The system will then fall near the bottom of
this adjacent well and the process will repeat. In this regime, the
dynamics can be reduced to a basin hoping: the local minima of the
energy become the nodes on the network, two such nodes are connected
by an edge if the system can transit from one minimum to the another
by crossing a single barrier, and the rate/weight of the directed edge
from one node to another involves (via Arhennius formula) the height
of the energy barrier(s) that must be crossed to perform this
transition -- this construction was illustrated on a toy example in
Fig.~\ref{fig:7well}. An additional simplification made in the case of
$\lj38$ is to lump together all the minima and saddle point that are
equivalent by symmetry (point group, permutation, etc.). All together
this construction led to a network for $\lj38$ that contains a single
connected component with 71887 nodes associated with the lowest local
minima on the landscape (which include FCC and ICO), and 119853 edges
-- this information is publicly available from the $\lj38$ database
Wales's website~\cite{wales_network}. The database also contains the
information about the generator, whose off-diagonal entries are in a
form consistent with~\eqref{eq:13}~\cite{wales1}
\begin{equation}
  \label{eq:1}
 L_{i,j} = \sum_{k}\frac{O_i\bar{\nu}_{i}^\kappa}
 {O^k_{i,j}(\bar{\nu}^k_{i,j})^{\kappa-1} }e^{-\beta (V^k_{i,j}-V_i)} 
\end{equation}
Here $\beta = 1/k_B T$ is proportional to the inverse of the system's
temperature $T$, $O_i$, $V_i$, and $\bar{\nu}_i$ are, respectively,
the point group order, the value of the potential energy, and the
geometric mean vibrational frequency for the local minimum associated
with node~$i$, $O^k_{i,j}=O^k_{j,i}$, $V^k_{i,j}=V^k_{j,i}$ and
$\bar{\nu}^k_{i,j}=\bar{\nu}^k_{j,i}$ are the same numbers for the
transition state $k$ connecting the local minima $i$ and $j$ (there
may be more than one of them for every pair $(i,j)$ adjacent on the
network), and $\kappa=3\times38-6=108$ is the number of vibrational
degrees of freedom. As in~\eqref{eq:13}, if there is no minimum energy
path connecting the minima with index $i$ and $j$ via a single saddle
point, we set $k=1$ and $V^{1}_{i,j}=\infty$. Note that, by
construction, the generator defined by~\eqref{eq:1} satisfies
detailed-balance with respect to the following Boltzmann-Gibbs
equilibrium distribution:
\begin{equation}
  \label{pi}
  \pi_i=\frac1{Z(\beta)}\frac{e^{-\beta V_i}}{O_i\bar{\nu}_{i}^\kappa}\qquad\qquad
  Z(\beta)=\sum_{i\in S} \frac{e^{-\beta V_i}}{O_i\bar{\nu}_{i}^\kappa}.
\end{equation}

The network representation of~$\lj38$ via~\eqref{eq:1} will be our
starting point here. The majority of local minima/nodes listed in
Wales database do not have special names -- for example, FCC and ICO
are simply listed 1st and 7th, respectively. Except for these two, in
the sequel we will simply refer to the other minima by their indices
in the database. We also work in reduced units in which the
temperature is measured in units of $a/k_B$. Since we are
interested in the mechanism of rearrangement between ICO and FCC, we
take the nodes of these two states as sets $A$ and $B$,
respectively. We also checked that our results do not change
significantly if we fatten these states by including in them the nodes
that are in the connected component around them where all the nodes
have energy within $k_BT$ of that of FCC and ICO, respectively.

\subsection{Computational Aspects of the TPT Analysis of $\lj38$}
\label{sec:comp}

A key preliminary step in the application of TPT to $\lj38$ is the
calculation of the committor function. This calculation requires
solving \eqref{lq} which, in the present case, is a system of
$71887-2=71885$ linear equations with the same number of unknowns.
The detailed balance property~\eqref{detbal} allows us to make the
matrix in \eqref{lq} symmetric by multiplying each row by
$(O_i\bar{\nu}_i^{\kappa})^{-1}e^{-\beta V_i}$. The resulting system
can then be solved using the conjugate gradient method with the
incomplete Cholesky preconditioning (see e.g.~\cite{CG}). This works
for $T\ge 0.125$.  For lower values of the temperature, the scale
separation between the possible values of $e^{-\beta V_i}$ for
different~$i$ becomes too large for the computer arithmetics.  In
order to overcome this difficulty we truncate the $\lj38$ network by
keeping only the nodes whose energy is below a given cap -- this is
legitimate because, the lower the temperature, the least likely it is
to observe a reactive trajectory venturing at energies much higher
than $k_BT$ above that of the overall barrier between ICO and FCC.
For each value of temperature we set this cap as high as possible
while keeping the system nonsingular in the computer arithmetics.  The
energy caps and resulting network sizes for the different values of
temperature that we considered are listed in Table~\ref{table1}.  The
values in parentheses are the difference between the caping energy and
that of FCC, $V_{{\rm FCC}}=-173.928$ \cite{wales_network}. All in
all, we computed the committor for temperatures ranging from $T=0.04$
to $T=0.18$ using steps of $\Delta T = 0.005$.

\begin{figure}[t]
  \centerline{\includegraphics[width=1\textwidth]{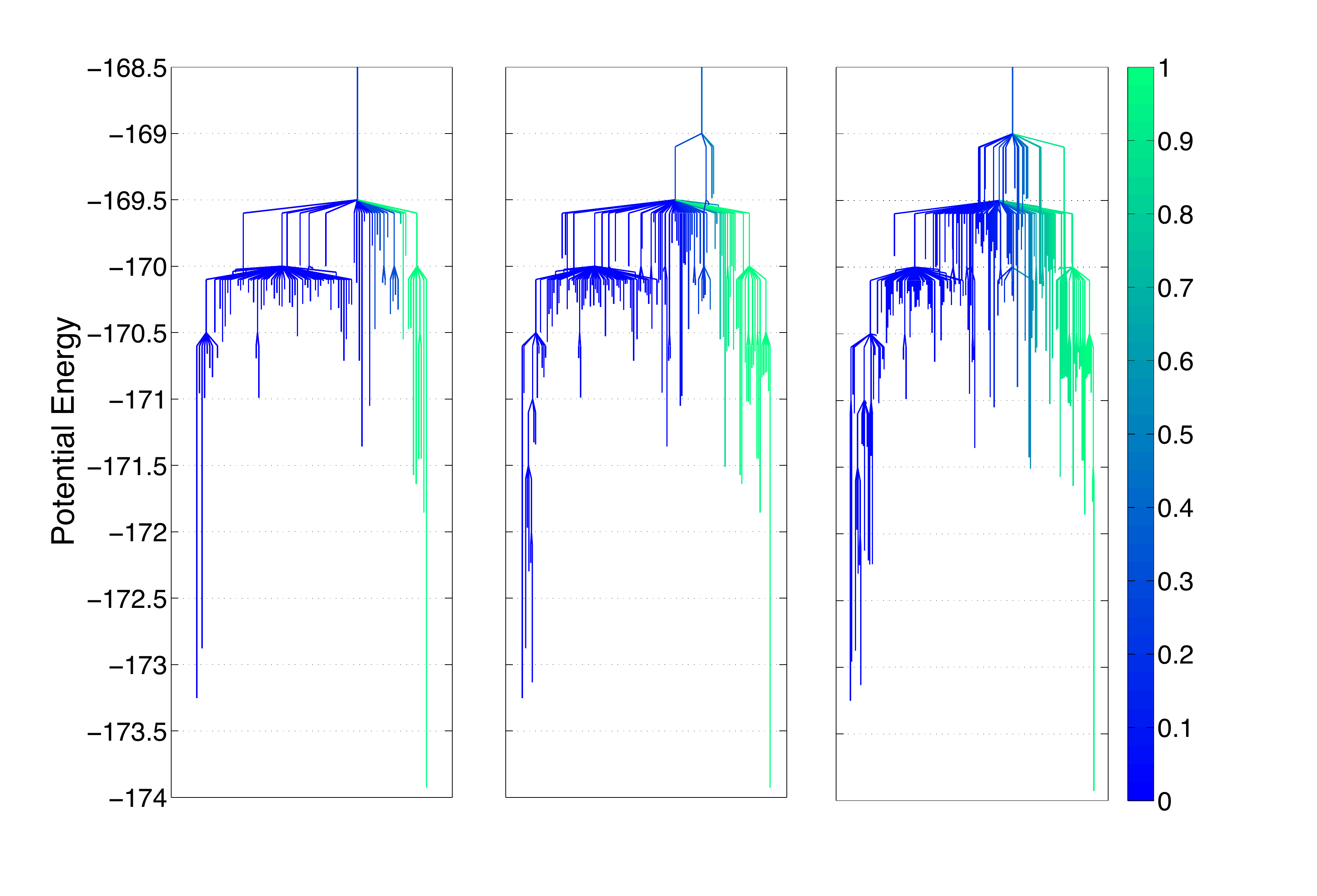}}
  \caption{Disconnectivity graphs colored according to the value of
    the committor: $T=0.06$ (left), $T=0.12$ (center), $T=0.15$
    (right).  Each disconnectivity graph includes only those local
    minima through which at least 1\% of the reaction pathways from
    ICO to FCC pass.}
  \label{fig:dgraph}
\end{figure}

The disconnectivity graphs of the network we used at three different
temperatures, $T= 0.06$, $T=0.12$, and $T=0.15$ are shown in
Fig.~\ref{fig:dgraph}. On these figures, we only included the nodes
through which at $1\%$ of the total current of reactive trajectories
goes and we colored the branches of the graph according to the value
of the committor of the nodes at the end of these branches. As can be
seen, as the temperature increases, the committor function becomes
less step-like, and a higher number of nodes gets values than are in
between the extreme 0 and 1.

\begin{table}[htdp]
  \caption{Energy caps and network sizes used for different value of the temperature.}
  \begin{center}
    \begin{tabular}{|c|c|c|}
      Temperature $T$ & Energy cap &Number of states \\
      \hline
      0.04 $\le T\le$ 0.05 & -169.5 (4.428) & 1604 \\
      0.055 & -168.5 (5.428) & 15056 \\
      0.06,  0.065 & -168.0 (5.928) & 28486 \\
      0.07 & -167.0 (6.928) & 53566 \\
      0.075, 0.08 & -166.5 (7.428) &  61706 \\
      0.085 & -165.5 (8.428) & 69302 \\
      0.09, 0.095 & -165.0 (8.928) & 70552 \\
      0.10 $\le T\le $ 0.12 & -164.0 (9.928) & 71609\\
      0.125 $\le T \le  0.18$ & $\infty$ & 71887\\
    \end{tabular}
  \end{center}
  \label{table1}
\end{table}

\begin{figure}[t]
  \centerline{\includegraphics[width=1\textwidth]{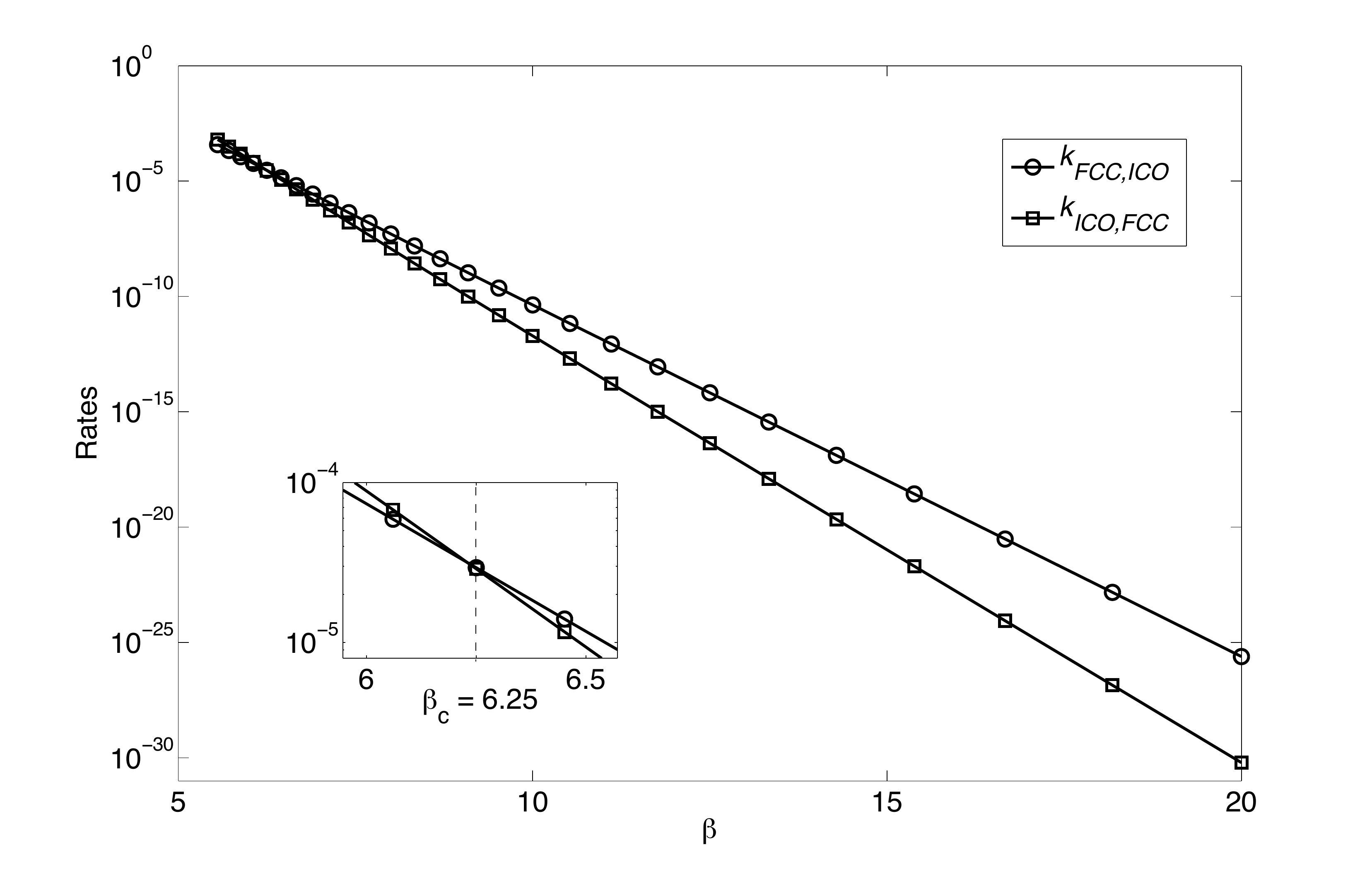}}
  \caption{The reaction rates in~\eqref{rel} computed with $A=$ ICO
    and $B =$ FCC at different temperatures. These rates display an
    almost perfect Arrhenius-like behavior in this temperature range,
    even though the mechanism of the rearrangement becomes increasingly
    complex as the temperature increases. The zoom shown in the inset
    shows that a cross-over between $k_{\text{FCC,ICO}}$ and
    $k_{\text{ICO,FCC}}$ occurs ar $\beta_c = 6.25$ (i.e. $T_c= 0.16$):
    this is the temperature above which ICO becomes more favorable than
    FCC due to entropic effects related to the relative widths of the
    funnels around these two structures.}
  \label{fig:rate}
\end{figure}

\begin{figure}[t]
\begin{center}
\centerline{
{\includegraphics[width=1\textwidth]{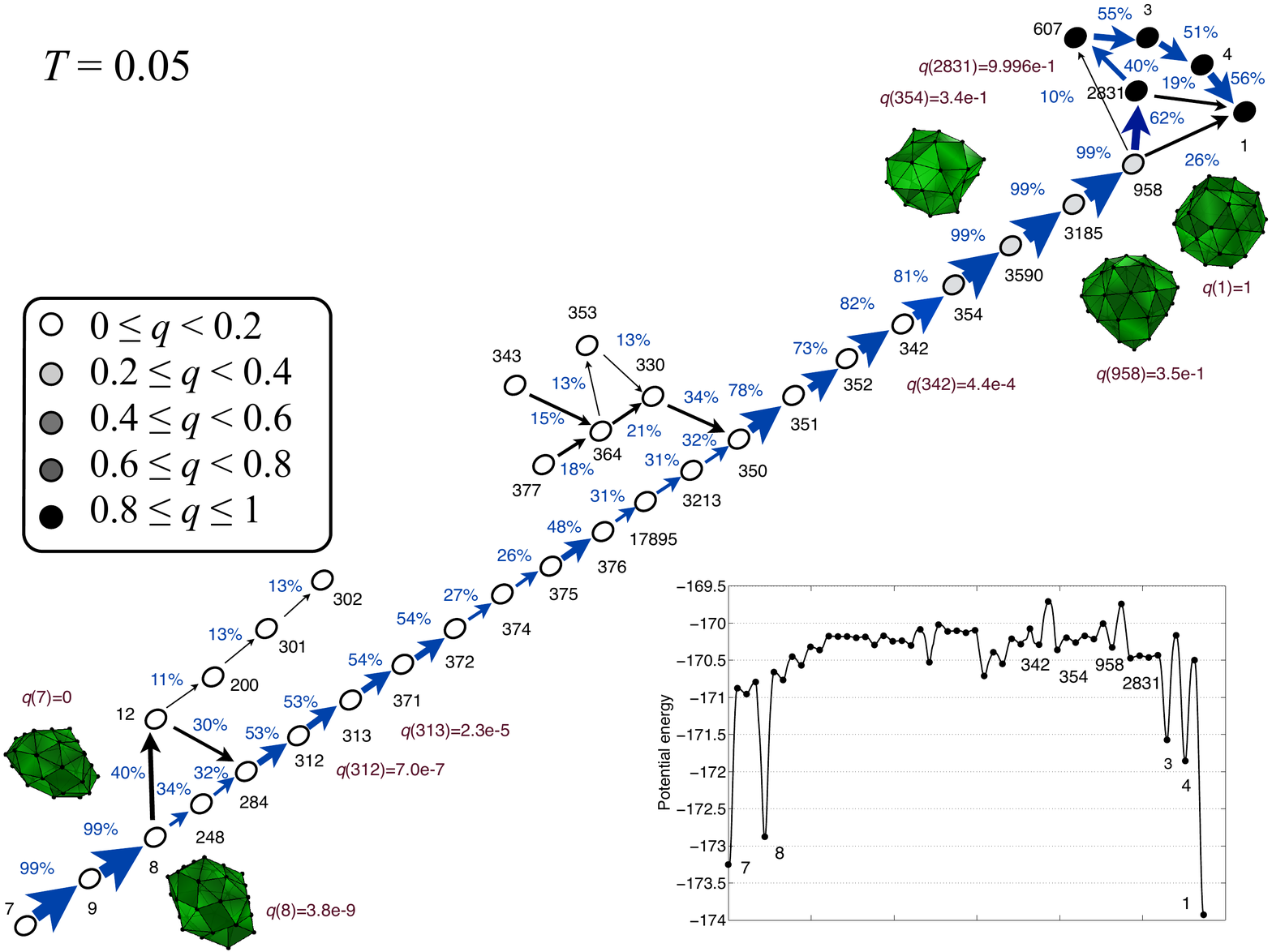}}
}
\caption{Cartoon representation of the network of current of reactive
  trajectories at $T=0.05$. The edges shown carry at least 10\% of the
  total reactive flux from ICO to FCC -- the thickness of the arrow is
  proportional to the precentage of current the edge carries, and the
  actual percentage is also displayed next to it.  The values of the
  committor at the nodes are show in greyscale, with the explicit
  values of $q$ given for some of them.  The blue arrows show the
  minmax path from LDT: at this low temperature, most of the current
  goes along this path. The highest barrier crossed along the minmax
  path is between nodes $342$ and $254$ ($V_{(342,254)}=4.219$). Also
  show in inset is the energy profile along the minmax path.}
\label{fig:cartoon1}
\end{center}
\end{figure}

\begin{figure}[t]
\begin{center}
\centerline{
{\includegraphics[width=\textwidth]{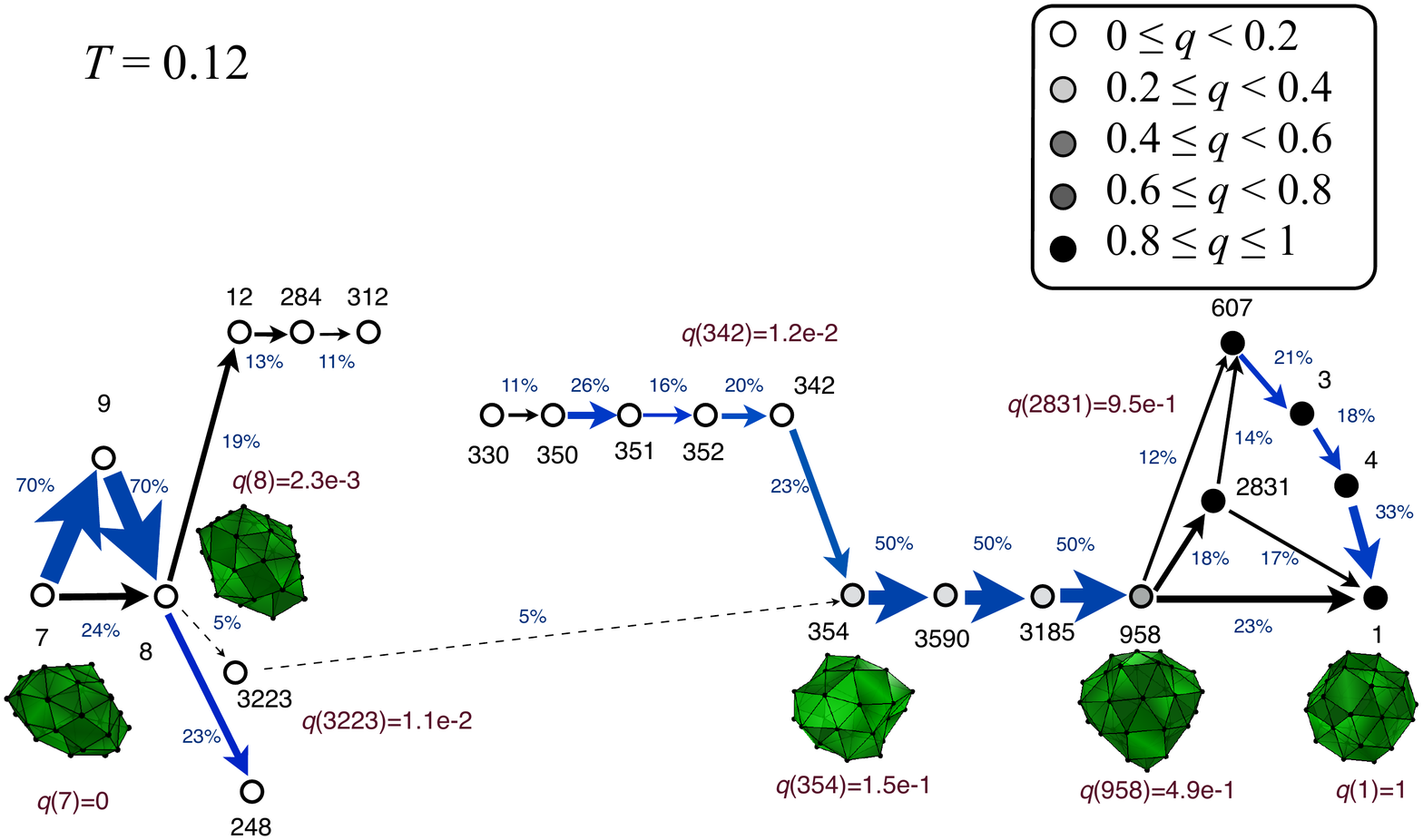}}
}
\caption{ Same as in Fig.~\ref{fig:cartoon1} at $T=0.12$. At this
  higher temperature, most of the edges carry less than $10\%$ of the
  total current: in particular, we can no longer go from ICO to FCC by
  following edges that carry at least $10\%$ of the current. This also
  implies that the minmax path from LDT is no longer relevant to
  explain the mechanism of the rearrangement at this temperature --
  the edges along this path that carry more than $10\%$ of the current
  are still shown in blue. The edges between nodes 8 and 3223 and
  nodes 3223 and 354 carry less than $5\%$ of the current: we show
  them because these edges belong to the dominant representative path
  introduced, i.e. the path maximizes the current it carries. This
  path is different from the minmax path but, as can be seen in this
  example, it is not relevant either in situations where the
  transition channel becomes spread out. }
\label{fig:cartoon2}
\end{center}
\end{figure}

\begin{figure}[t]
  \centerline{\includegraphics[width=1\textwidth]{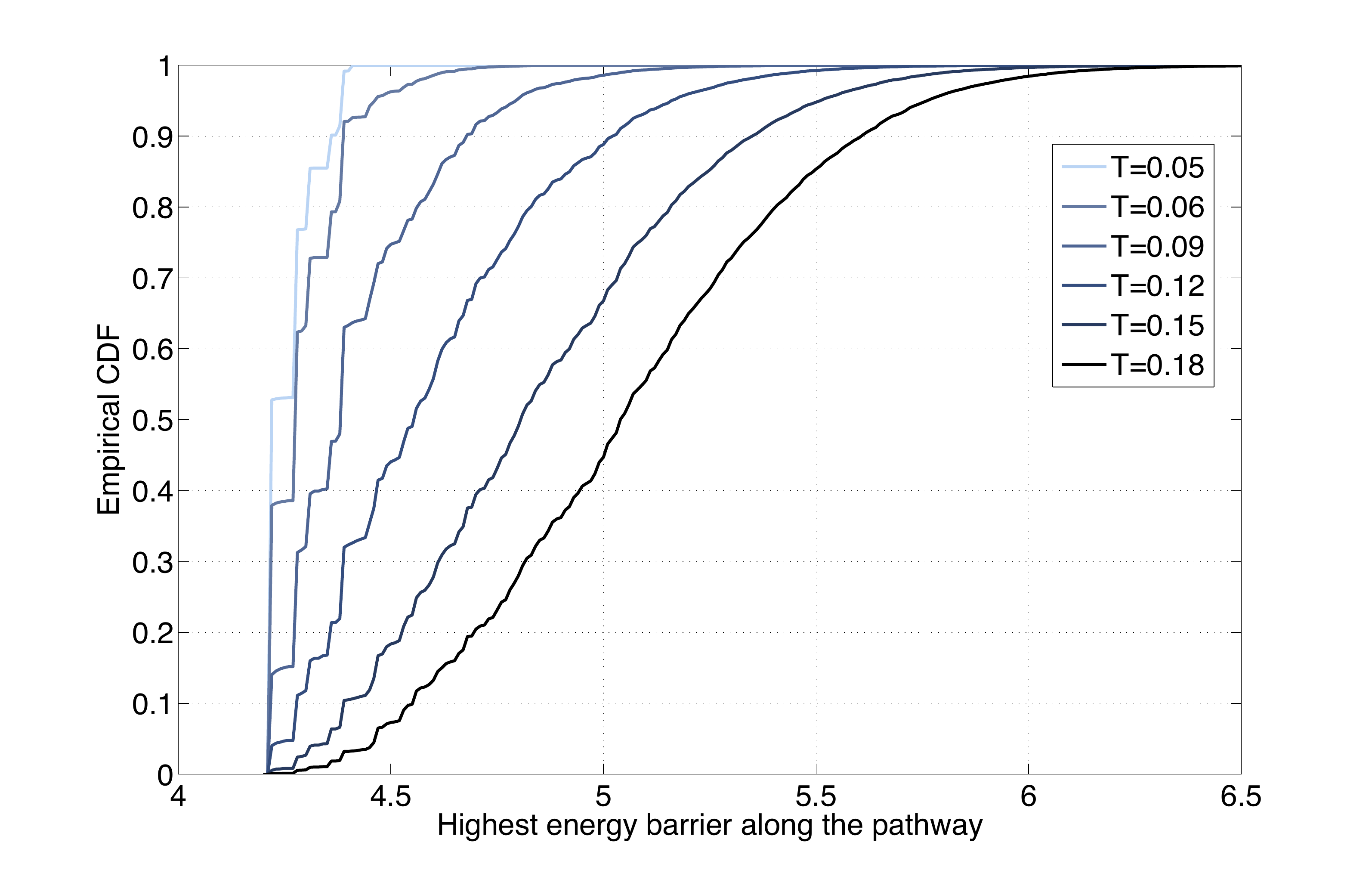}}
  \caption{Empirical cumulative distribution function of the highest
    energy barrier along the no-detour paths at different temperature.
    The height of the barriers are given respective to the energy of
    FCC, $V_{\text{FCC}}=−173.928$. As the temperature increases, the
    no-detour paths tend to cross higher barriers. }
  \label{fig:barriers}
\end{figure}

\begin{figure}[t]
  \centerline{\includegraphics[width=1\textwidth]{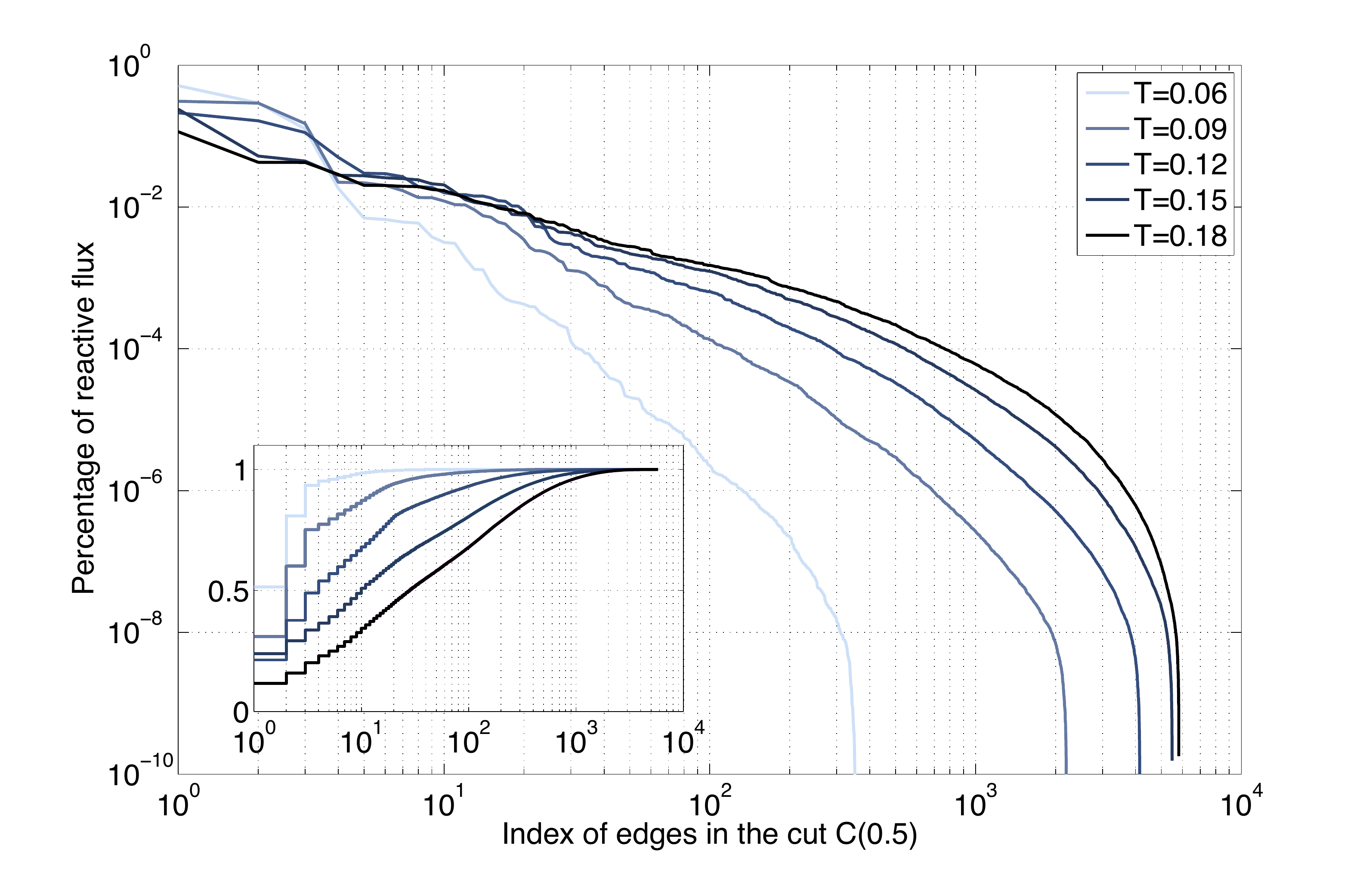}}
  \caption{ The magnitude of the current through the edges in the
    isocommittor cut $C(0.5)$ is plotted against the index of these
    edges ordered by this magnitude. The inset shows the empirical
    cumulative distribution function of the current through the edges
    in the cut.  The number of edges that must be included to account
    for a given percentage of the total current increases rapidly with
    temperature, indicative of the broadening of the reaction channel
    for the rearrangement.}
  \label{fig:fluxincut}
\end{figure}

\begin{figure}[t]
  \centerline{\includegraphics[width=1\textwidth]{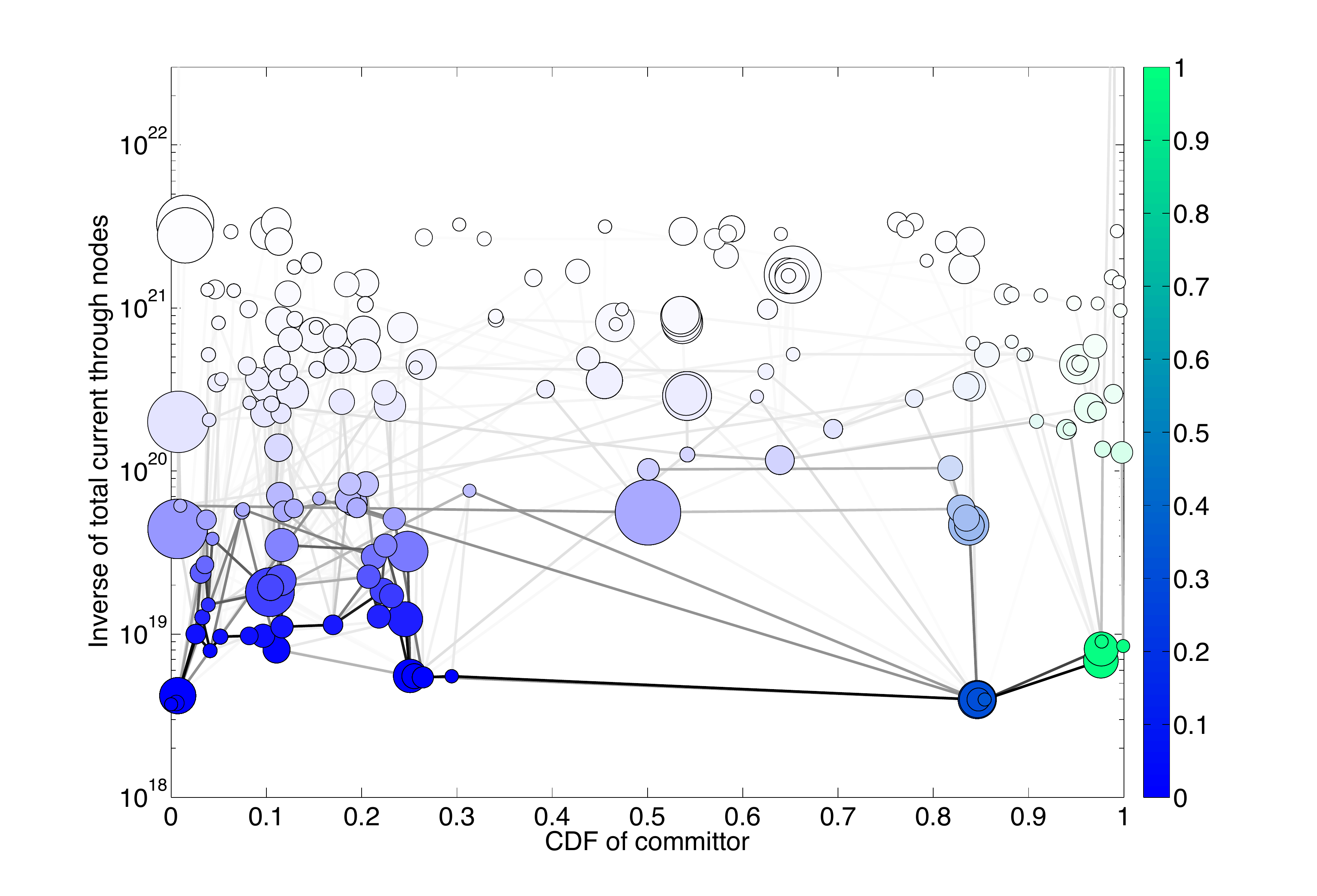}}
  \caption{Graphical representation of the network of reactive current
    in $\lj38$ at $T=0.06$. The way this representation was constructed
    is explained in text.}
  \label{fig:net06}
\end{figure}

\begin{figure}[t]
  \centerline{\includegraphics[width=1\textwidth]{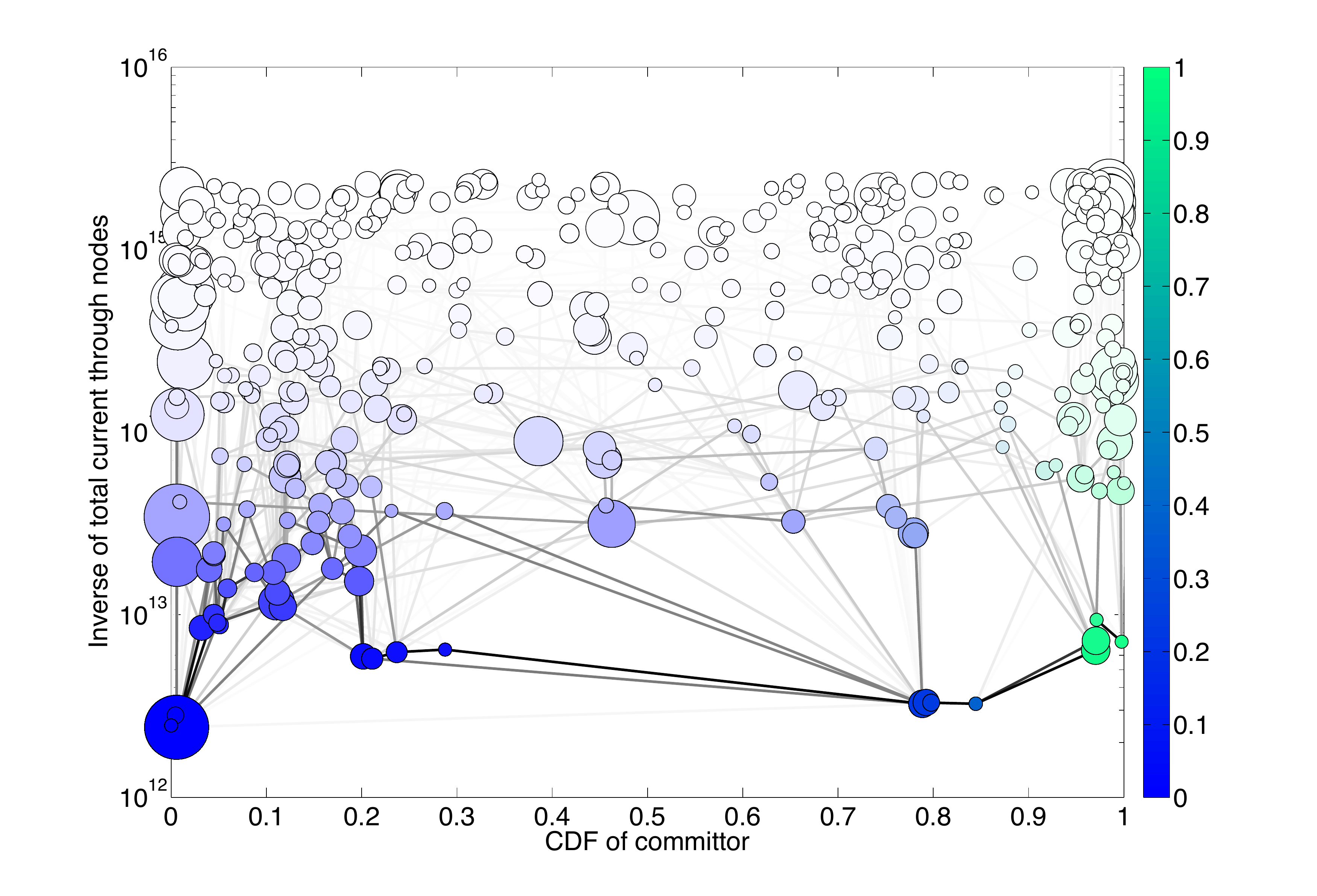}}
  \caption{Same as in Fig.~\ref{fig:net06} at $T=0.09$. }
  \label{fig:net09}
\end{figure}

\begin{figure}[t]
  \centerline{\includegraphics[width=1\textwidth]{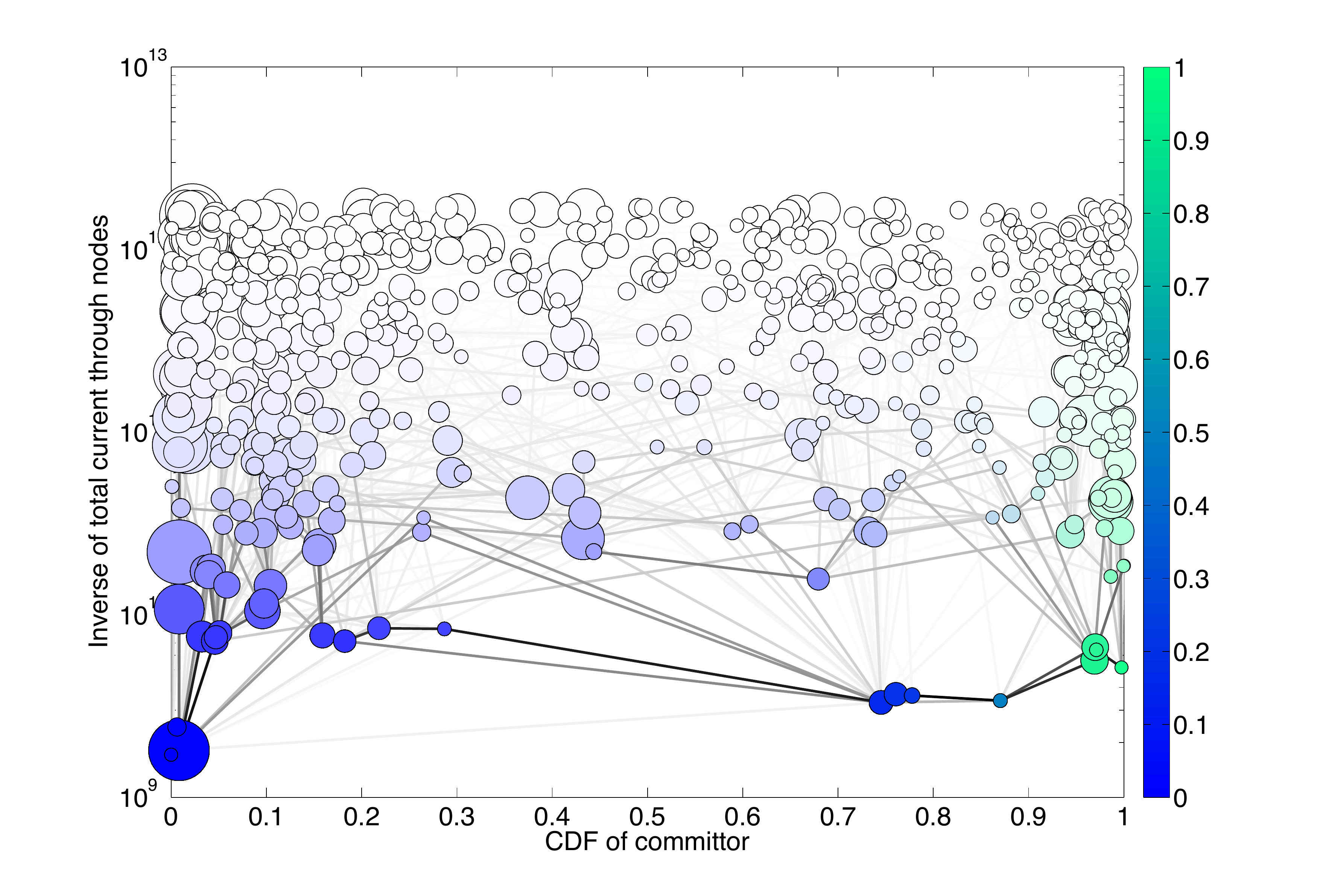}}
  \caption{Same as in Fig.~\ref{fig:net06} at $T=0.12$.}
  \label{fig:net12}
\end{figure}

\begin{figure}[t]
  \centerline{\includegraphics[width=1\textwidth]{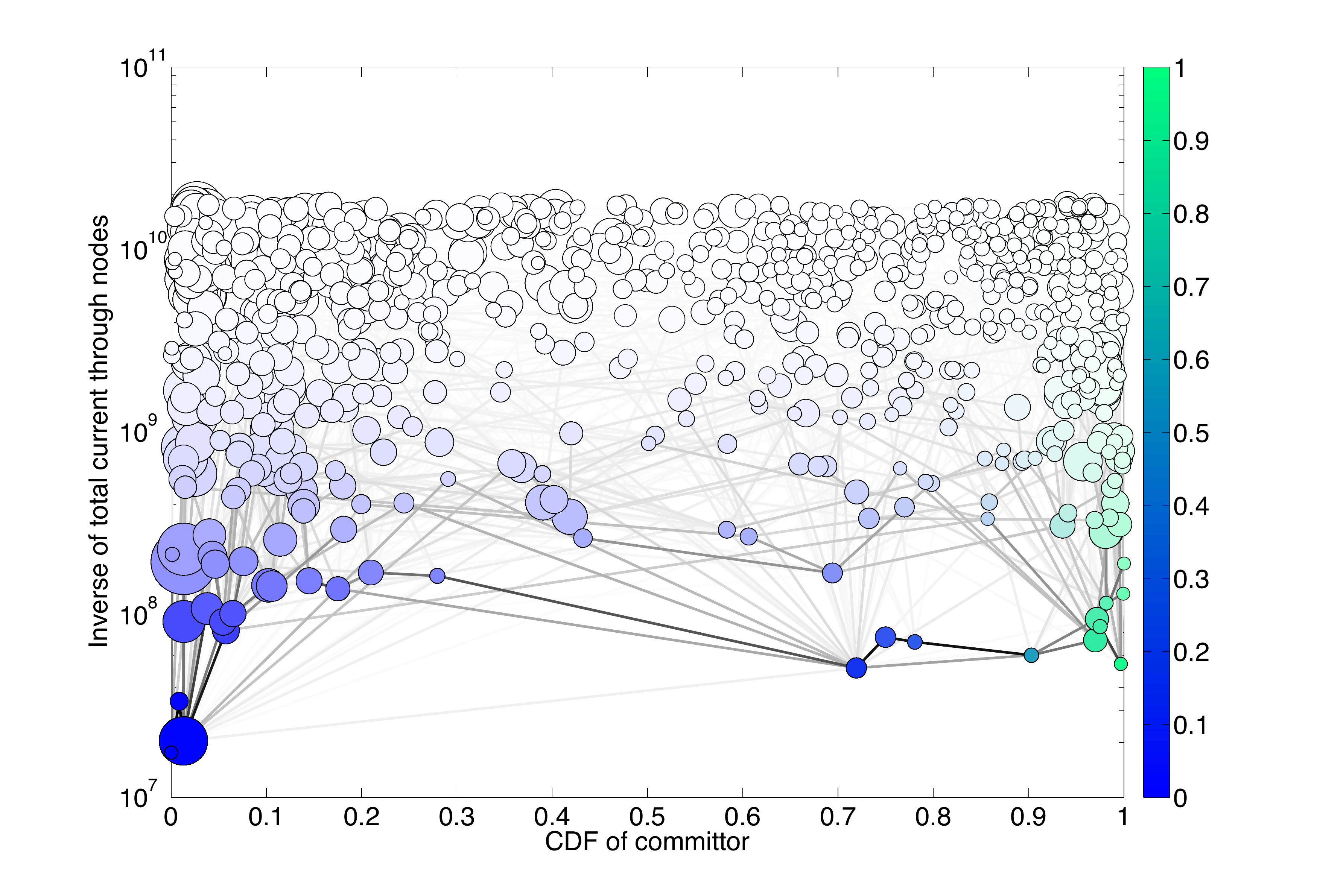}}
  \caption{Same as in Fig.~\ref{fig:net06} at $T=0.15$.}
  \label{fig:net15}
\end{figure}

\subsection{Rate and Mechanism of Rearrangement at Different
  Temperatures}
\label{sec:cartoon}

Once the committor function has been calculated, we can use TPT to
calculate the rate of rearrangement of $\lj38$ and characterize its
mechanism. Using formulae~\eqref{rel} with $A=$ ICO and $B=$ FCC, we
obtain the rates at which the system rearranges itself between these
two states. These rates are shown in Fig.~\ref{fig:rate} as a function
of the inverse temperature~$\beta$. As can be seen, both rates are
almost perfectly straight on a log-linear scale, and can be fitted by
\begin{equation}
  \label{eq:17}
  k_{\text{FCC,ICO}} = 1.03 \times 10^7 e^{-4.289 \beta}, \qquad 
  k_{\text{ICO,FCC}} = 9.81 \times 10^4 e^{-3.525 \beta}
\end{equation}
Since the energy barriers between FCC and ICO and ICO and FCC are
4.219 and 3.543, respectively, \cite{wales38}, these fits are
consistent with Arrhenius law.  The fits in~\eqref{eq:17} also compare
well with the ones calculated in \cite{wales_landscapes} for the
temperature range $0.03\le T\le 0.4$:
$k_{\text{FCC,ICO}}=2.13\times10^6e^{-4.29\beta }$ and
$k_{\text{ICO,FCC}}=1.16\times10^3e^{-3.43\beta}$. Note also that the
rates cross at the value $\beta = 6.25$ (i.e. $T= 0.16$): this
temperature is the one above which TPT predicts that ICO becomes
preferred over FCC, which is slightly higher than the value $T_c=0.12$
listed in Sec.~\ref{sec:thermo}. This crossover is due to entropic
effects related to the relative widths of the funnels around ICO and FCC.

The Arrhenius-like nature of the rates may suggest that the mechanism
of rearrangement of the $\lj38$ cluster is quite simple, and dominated
at all the temperatures that we considered by the hoping over the
lowest saddle point separating ICO and FCC. This impression, however,
is deceptive. To see why, in Figs.~\ref{fig:cartoon1}
and~\ref{fig:cartoon2} let us compare cartoon representations of the
current of reactive trajectories given in~\eqref{eq:effcurrent} at two
different temperatures, $T=0.05$ and $T=0.12$. The way these
representations were constructed is by plotting all the nodes in the
network such that the current of reactive trajectories along the edges
between them carry at least 10$\%$ of the total current, and
connecting these nodes by an arrow whose thickness is proportional to
the magnitude of the current. As can be seen in
Fig.~\ref{fig:cartoon1}, at $T=0.05$, most of the current concentrate
on a single path: this path coincides with the minmax path between ICO
and FCC predicted by LDT~\cite{cam1}. At the higher temperature of
$T=0.12$, however, we see that this minmax path becomes mostly
irrelevant, and in fact we can no longer go from ICO to FCC following
edges that carry at least 10$\%$ of the current. The reason is that
the current becomes very spread out among the edges of the network,
indicative that the tube carrying most of the current of reactive
trajectories also becomes quite wide.

To quantify further this observation, we used
Proposition~\ref{th:lftpp} to generate $10^8$ samples of the no-detour
transition path process at every temperature. (In the present example,
it turns out that the network is so complex that the reactive
trajectories themselves, which we can in principle generate via
Proposition~\ref{th:tpp}, are too long to be sampled efficiently. This
arises because these trajectories wander too often into quasi-deadends
or in between intermediate structures, and this is why we focused on
no-detour transition paths which are much shorter and can be generated
in great number.) We used this sample of no-detour transition paths to
first analyze the height of the highest energy barrier along these
paths measured with respect to $V_{\text{FCC}}= - 173.928$.  The
empirical cumulative distribution functions of these barrier heights
are shown in Fig.~\ref{fig:barriers}. As can be seen, at the low
temperature of $T=0.05$, this distribution is very peaked around the
value $4.219$, which is the height of the lowest saddle point
separating ICO and FCC. At higher temperatures, however, this
distribution broadens significantly, indicative that higher barriers
become frequently crossed by the no-detour transition paths. This is
an entropic effect: in essence, we can think of the height of the
barrier in terms of `bonds' between the Lennard-Jones particles that
need to be broken for the rearrangement to proceed. What our results
show is that the number of no-detour paths increases very rapidly with
the maximal number of bonds that are ever broken along them. At low
temperature, the rearrangement proceed mostly by no-detour paths along
which no more than about 4 bounds are broken, because these paths are
energetically favorable. At higher temperature, however, no-detour
paths along which 5, 6 or even 7 bonds break start to matter: even
though they are less favorable energetically, their sheer number
means that they eventually carry more current globally.

A consequence of this effect is that the width of the reaction channel
also broadens significantly with temperature.  This is quantified in
Fig.~\ref{fig:fluxincut}, where we analyze the current along the edges
in the isocommittor cut $C(0.5)$. By ordering these edges by the
magnitude of the current they carry, and plotting this current magnitude
as a function of the edge index, we arrive at the plots on the main
panel of Fig.~\ref{fig:fluxincut}. As can be seen, as the temperature
increases, these plots widen with temperature, and display a power
law behavior for a range of edge indices. The inset of
Fig.~\ref{fig:fluxincut} shows the cumulative distribution of the
current through the edges in the isocommittor cut $C(0.5)$, and show
that the higher the temperature, the more edges need to be included to
get a significant percentage of the total current: for example, at
$T=0.18$, thousands of edges in the cut (that is, most of them) need
to be included in order to account for $95\%$ of the current. The
mechanism of rearrangement thus departs significantly from the one
predicted by LDT, even though the rates remain Arrhenius-like even at
this high temperature.

We tried to capture visually the complexity of the mechanism of
rearrangement using the representation of the network of current of
reactive trajectories shown in Figs.~\ref{fig:net06}--\ref{fig:net15}.
These figures were constructed as follows. We plotted every node of
the network through which at least $0.1\%$ of the total current
went. We ordered these nodes along the $x$-axis according to the
cumulative distribution function of their committor, using a coloring
from blue to green to indicate their actual committor value. Along the
$y$-axis, we ordered the nodes according to the inverse of the
magnitude of current of reactive trajectory, \eqref{eq:effcurrent},
they carry (the higher the node, the least current it carries) and we
connected the nodes by lines whose darkness is proportional to the
magnitude of the current between them. We also faded the color as this
magnitude decreased. Finally, we used dots of different sizes to
represent the nodes: the bigger the node, the larger is the magnitude
of the average number of transitions per unit time that the reactive
trajectories make through this node, see \eqref{dpc}. This is a way to
try to capture deadends and dynamical traps on the network, i.e. node
that the reactive trajectories visit often but through which little
current of these reactive trajectories go. In the figures these
deadends are nodes that are high and big. Overall, what these figures
confirm is that, as the temperature increases, the curent of reactive
trajectories spreads more and more on the network, and the reaction
channel broadens. It also confirms that there exists many deadends and
dynamical traps on the network. This last aspect makes TPT
particularly suitable to analyze the mechanism of rearrangement:
indeed, a spectral analysis of the network along the lines discussed
in Sec.~\ref{sec:spectral} is both hard to perform in the present
situation and uninformative because it is too global.

\section{Outlook and Conclusions}
\label{sec:conclu}

We have presented a set of analytical and computational tools based on
TPT to analyze flows on complex networks/MJPs. We expect these tools
to be useful in a wide variety of contexts.  The network
representation of $\lj38$ that we used here as illustration is just a
specific example of Markov State Model (MSM) used to map a complex
dynamical system onto a MJP (see e.g.~\cite{MSM}). During the last
decade, such MSMs have emerged as a way to analyze timeseries data
generated e.g. by molecular dynamics simulations of macromolecules,
general circulation models of the atmosphere/ocean system, etc. In
these contexts, massively parallel simulations, special-purpose
supercomputers, and high-performance graphic processing units (GPUs)
permit to generate time series data in amounts too large to be grasped
by traditional ``look and see'' techniques. MSMs provide a way to
analyze these data by partitioning the conformation space of the
molecular system into discrete substates, and reducing the original
kinetics of the system to Markov jumps between these states -- in
other words, by interpreting the timeseries as some dynamics on a
network, with the states in the MSMs playing the role of the nodes on
the network, and the transition rates between these states being the
weights of the directed edges between these nodes.  While MSMs
typically provide an enormous simplification of the original
timeseries data, the associated networks are typically quite complex
themselves, with many nodes, a nontrivial topology of edges between
them, and rates/weights on these edges that can span a wide range of
scales. The tools that we derived from TPT can be used for the
nontrivial task of analyzing these networks/MSMs.

More generally, we expect the tools developed in this paper to be
useful to analyze and interpret other networks that have emerged in
many areas as a way to represent complex data sets.

\section*{Acknowledgments} We thank Prof. David Wales for providing us
with the data of the $\lj38$ network and Miranda Holmes for
interesting discussions.  M. C. held an Sloan Research Fellowship and
was supported in part by DARPA YFA Grant N66001-12-1-4220, and NSF
grant 1217118. E. V.-E.  was supported in part by NSF grant
DMS07-08140 and ONR grant N00014-11-1-0345.



\bibliographystyle{spphys}       
\end{document}